\documentclass{aa}  

\usepackage{graphicx}
\usepackage{txfonts}
\usepackage{lipsum}
\usepackage{subcaption}         
\usepackage{lscape}             
\usepackage{placeins}           
                                
\begin{document}

   \title{Challenging historical novae: \\ AT Cnc (1645), Te-11 (483), and M22 (BC 48) revisited}

   \titlerunning{Historical novae AT Cnc (1645), Te-11 (483), and M22 (BC 48)}
   

   \author{D.L. Neuh\"auser\inst{1}
        \and R. Neuh\"auser\inst{2}\fnmsep\thanks{corresponding author: ralph.neuhaeuser@uni-jena.de}
        \and V. Hambaryan\inst{2,3,4}
        \and J. Chapman\inst{5}        
        \and M. Della Valle\inst{6}
        }

   \institute{Independent scholar, 30126 Venezia, Italy 
            \and Astrophysical Institute, Friedrich-Schiller-University Jena, Schillerg\"a\ss chen 2, 07745 Jena, Germany 
            \and Byurakan Astrophysical Observatory after V.A. Ambartsumian, 0213, Byurakan, Aragatzotn, Armenia
            \and Astrophysical Research Laboratory of Physics Institute, Yerevan State University, Yerevan, Armenia
            \and Assistant Editor for Early China, Society for the Study of Early China, 1995 University Ave 510, Berkeley, CA 94706, USA
            \and INAF Napoli, Osservatorio Astronomico di Capodimonte, Salita Moiariello 16, 80131 Napoli, Italy}
   \date{19 March 2025 / 22 Aug 2025}
 
  \abstract
   {Connections between novae with shells and historical observations are crucial for 
   astrophysical understanding of long-term evolution of shells and cataclysmic variables.}
   {Three of five previously considered links are revisited here: extended features 
   in
   M22 in BC48, Te-11 in 483, and AT Cnc in 1645. 
   We aim to develop a procedure to check whether these links are credible.}
   {Literal translations of the Chinese texts, 
   historically based 
   arguments, and close readings are combined with astrophysics, 
   (peak brightness, decay time estimate, shell age expansion model calculation, etc.).}
   {(a) Nandou's 
   second 
   star, near which the BC48 `guest star' was reported, is identified as $\tau$ Sgr, not $\lambda$ Sgr, far from the M22 location. 
   A nova in M22 would peak at only m=6.4$\pm$1.4\,mag, 
   and thus 
   a description as 
   a
   `blue-white' `melon' does not fit; it was likely a comet. 
   (b) The imprecise position (`Shen['s] east') of the `guest star' in 483, its extended (dipper-like) radiance, and the context speak for a bolide. 
Considering the new (larger) Gaia distance and small extinction towards Te-11 (outside a cloud), 
its bi-polar morphology and current expansion velocity point to a 
planetary nebula; as a nova, the shell expansion 
age is $\sim$1100-2000\,yr from detailed supersonic expansion calculations.
(c) Most certainly, Mars was meant when the source for 1645 reported `a large star entered Yugui'; the verb implies motion.
AT Cnc lies neither in Yugui's 
asterism box 
nor in the eponymous 
lunar mansion 
range. 
The fluid drag expansion age of AT Cancri's ejecta is $\sim$128-631 yr.}
   {All three exact ages are 
   unsubstantiated.
True novae or nova shells can be connected to historical records only if 
the
position and object type are plausible.
Duration, brightness 
(light curve), 
and color (evolution) should fit and could provide more astrophysical insight.
Then, shell ages are sufficiently precise for properly 
studying secular evolution of novae, shell sizes, H$\alpha$ luminosities, long-term decay, etc.}


   \maketitle


\nolinenumbers

\section{Introduction}

Cataclysmic variables (CVs) are binary systems, in which
a low-mass star transfers material to a 
white dwarf 
(WD) companion. 
Instabilities in the accretion disk can lead to brightenings of up to 
approximately a hundred-fold
for days to months 
due to conversion of gravitational energy into heat -- these are called 
dwarf novae. Classical novae 
result from thermonuclear runaways of accreted hydrogen onto the surface of the WD. 
These events are much rarer, brighter, and can increase the luminosity up to $10^{5}$ times; they eject
nova shells, while the WD survives. White 
dwarfs 
in nova systems are thought to spend most of their time in 
hibernation between eruptions (Shara et al. 1986), with recurrence times of $10^{5-6}$ yr (Yaron et al. 2005) or more. 
However, this scenario has faced criticism (e.g., Schaefer 2023) or has been proposed in milder versions 
(e.g., Della Valle \& Izzo 2020). Thermonuclear supernovae (SN Ia) are significantly brighter and result 
from CV systems, 
whereby
a massive WD, nearing the Chandrasekhar limit, is completely disrupted (review in Livio \& Mazzali 2018).
 
The oldest undisputed 
classical nova
with a still observable shell is \object{T Aur} (1891); 
earlier records of (naked-eye) sightings may exist.
Five such links were considered previously:
\object{Z Cam} in BC 77 (Shara et al. 2007, 2012a, 2024), in M22 in BC 48 (G\"ottgens et al. 2019), 
\object{Te-11} in AD 483 (Miszalski et al. 2016), Nova-Sco in 1437 (Shara et al. 2017b),
and \object{AT Cnc} in 1645 (Warner 2016, Shara et al. 2012b, 2017a).
The cases in BC 48, AD 483 (considered also as 
planetary nebulae (PNe)), 
and AD 1645 are revisited here; 
discussions of the others will be published later.
Guerrero et al. (2025) recently found that \object{H$\alpha$ Tr 5} is a 
PN,
and therefore not related to Nova-Sco 1437.

Reliable historical novae provide essential input for astrophysics by offering insights into, 
for example,
the evolution of CVs, 
the behaviour of WDs, and the chemical enrichment of the Galaxy. Studying old novae helps refine models of stellar evolution, 
recurrence rates, and explosive phenomena, improving our understanding of parent stellar populations and Galactic dynamics: \\
(1) Evolutionary scenarios of close binaries (e.g. CVs), 
including the hibernation theory, expect a certain number of nova eruptions 
over a given period of time for a certain dwarf nova population size.
Therefore, the pure frequency of historical novae is of high relevance for nova rates and population studies 
(e.g., Shafter 2017, Della Valle \& Izzo 2020, Canbay et al. 2023). \\
(2) Hibernation theory predicts a 
brightness decay of 12 mmag yr$^{-1}$ for about a millennium (Kovetz et al. 1988); however, see Schaefer (2023).
Duerbeck (1992) determined a mean decay of 10$\pm$3 mmag yr$^{-1}$ from nine old novae,
several of which were observed in the early 20th century. \\
(3) Duerbeck (1987) estimated that the expansion velocity halves in 77$\pm$27 yr (free expansion)
from four early 20th century nova shells
(GK Per 1901, V603 Aql 1918, V476 Cyg 1920, DQ Her 1934).
However, Downes \& Duerbeck (2000, for HR Del from 1967) and
Santamaria et al. (2020, for T Aur 1891, V476 Cyg 1920, DQ Her 1934, V533 Her 1963, and FH Ser 1970)
found that there is no evidence 
of
deceleration. This would reduce the total lifetime of nova shells. \\
(4) The secular evolution of the H$\alpha$ luminosity of nova shells was recently studied by Tappert et al. (2020);
after 
around
150 yr, it remains constant around $10^{30}$ erg s$^{-1}$.
However, for the oldest shell-like features (including Te-11 and AT Cnc),
the age determination, or even the identification as nova, may be dubious. \\ 
(5) The study of recurrent novae, 
for example 
their death rate or repetition 
timescales 
(Schaefer 2010),
could be extended to 
timescales 
of centuries to millennia with historical novae. 

\begin{table*}
\begin{tabular}{llllcllccll} \hline
\multicolumn{11}{l}{Table 1: The three studied objects considered as historical novae with shell-like features:} \\ \hline
Name   & Nova & Object type\tablefootmark{(a)} & Mass WD & V range & \hspace{-.2cm} B-V & A$_{\rm V}$ & Distance     & \multicolumn{2}{c}{Nova\,peak\,[mag]} & Decay\tablefootmark{(d)}\\
      & year? & & [M$_{\odot}$] & [mag]\tablefootmark{(a)}& \hspace{-.2cm} mag & [mag] & d\,[pc]\tablefootmark{(b)} & \multicolumn{2}{l}{m~($\pm$1.4)~$\Delta$m\tablefootmark{(c)}} & dur. [yr] \\ \hline
in\,M22 & BC48 & \multicolumn{2}{l}{unknown (`blue-white')~~~...} & ... & ... & 1.0\tablefoottext{e} & \hspace{-.25cm} 3$\pm$0.3kpc\tablefoottext{f} & 6.4 & ~... & ... \\
Te-11  & 483   & ecl.UG,WD+M2.5 & $1.18^{+0.07}_{-0.15}$\tablefoottext{g} & 14.8-20.0 & 0.5 & 1.0\tablefoottext{g} & $526^{+48}_{-47}$\tablefootmark{(h)}& 2.6 & ~12.2-17.4 & 1480$\pm$530 \\ 
AT\,Cnc & 1645 & UGZ,DAe+K7/M0 & 0.9$\pm$0.4\tablefoottext{i} & 12.0-16.6 & 0.2 & 0.04\tablefoottext{j} & $455.7^{+7.1}_{-5.5}$ & 1.3 & ~10.7-15.3 & 1300$\pm$474 \\ \hline
\end{tabular}
\tablefoot{
\tablefoottext{a}{Object type and V range from VSX, Simbad, AAVSO, current quiescent color indices B-V from Simbad.} \\
\tablefoottext{b}{Bailer-Jones et al. (2021), photogeometric Gaia DR3 distance (except for M22).} \\
\tablefoottext{c}{Nova peak magnitude $m$ estimated from typical nova absolute magnitude M$_{\rm V}$=-$7.0\pm$1.4 mag 
(Schaefer 2018, Della Valle \& Izzo 2020), distance $d$, and extinction A$_{\rm V}$.
The calculated amplitude $\Delta$m is the difference between the expected peak and the quiescent V-band range;
$\Delta$m has to be ca.\,7-16 mag for novae (Warner 1995, figure 5.4) within $1\sigma$.
Both peak $m$ and amplitude $\Delta$m have the same uncertainty ($\pm 1.4$ mag). 
Note that Schaefer (2018) does not recommend using the relation between typical M and peak, at least not as first choice.
However, this reservation is meant for determining the distance from the observed peak, if a Gaia parallax is unavailable,
but in our historical novae, the peak magnitude is not known.} \\
\tablefoottext{d}{Duration from the estimated peak to the current quiescent magnitude with a mean decay of 
10$\pm$3 mmag yr$^{-1}$ (Duerbeck 1992).} \\
\tablefoottext{e}{Average from Samus et al. (1995), Monaco et al. (2004), 
and G\"ottgens et al. (2019); the latter neglected extinction in the peak estimate.} \\
\tablefoottext{f}{Monaco et al. (2004).} \\
\tablefoottext{g}{Miszalski et al. (2016).} \\
\tablefoottext{h}{The Gaia parallax quality indicator is 0.18704 and exceeds its recommended threshold of 0.1, 
probably due to the binarity, so that the distance may be more 
uncertain, as 
the error bars indicate.} \\
\tablefoottext{i}{Nogami et al. (1999).} \\
\tablefoottext{j}{Gaia A$_{\rm G}$.}
}
\end{table*}

A historical record of a classical nova can be used to date its shell far more precisely than any
astrophysical technique, and it may also constrain 
its speed class, light curve, and color evolution.
Shell expansion is studied here using 
a detailed three-phase physical model, which is
the best (or even only)
alternative for nova shell age determination other than
historical observations.

A homogeneous text corpus of transient celestial phenomena 
has been transmitted in 
classical 
Chinese for some two millennia.
For compilations of English translations, 
see for example 
Ho (1962), Xu et al. (2000), and Pankenier et al. (2008);
the latter two also include the original text in 
classical 
Chinese.
For reviews on historical Chinese astronomy, see Needham \& Wang (1959) and Sun \& Kistemaker (1997). 

To differentiate between 
(super-)novae  
and other transients,
such as
comets, bolides, or unrecognised planets,
it is important to study the context with all relevant information. 
Comets can be misinterpreted as (super-)novae, if motion relative to stars and extension due to coma or tail are not mentioned explicitly
(Neuh\"auser et al. 2021a);
see D.L. Neuh\"auser et al. (2021b) for comet criteria. 
The methods used here for close reading and text critique are summarized in 
Neuh\"auser et al. (2020) and Neuh\"auser \& Neuh\"auser (2021, therein section 3.1).  

Establishing a historical nova, one can go 
from history to astrophysics, i.e. searching for nova-like objects based on historical records
(e.g. Shara et al. (2017b) for the guest star of AD 1437, but see Guerrero et al. (2025)),
or alternatively from astrophysics to history, 
i.e. searching for historically observed transients, whose
position and object type fit a nova shell with its CV -- the latter 
would be the case for the three links discussed here.

In principle, historical sources of celestial observations and astrophysical findings and knowledge are 
independent. 
A well-fitting connection can yield unique insights into the 
the evolution of CVs, 
such as
the exact date of the classical nova 
explosion, and hence the shell age.
Both strategies involve the following methodological demands:\\
(a) the 
reported location of the transient has to be consistent with the position of the CV and its shell; \\
(b) the historical phenomenon must be
compatible with a classical nova, in particular star-like, i.e. stationary and non-extended; \\ 
(c) serendipitous discovery by the unaided eye, which typically needs an apparent peak of ca.\,2.5 mag or brighter
(e.g. Neuh\"auser \& Neuh\"auser 2021, therein section 2.4), has to be consistent with the given distance; \\
(d) the duration of observation has to be consistent with what is known about the nova candidates regarding distance 
(brightness) and mass (decline speed) -- a lack of a reported duration would weaken the credibility of the nova interpretation; and \\
(e) the precise historical date must not be inconsistent with the rough nova shell expansion age.

Set-up: 
This interdisciplinary study considers three suggested links between (presumable) novae with shells and historical observations: 
an extended feature in \object{M22} from BC 48 (Sect. 2), Te-11 as nova shell for a guest star of AD 483 (Sect. 3), and dwarf nova AT Cnc for an 
observation in AD 1645 (Sect. 4); see Table 1. 
In Sect. 5, model calculations of the supersonic expansion of nova shells are performed 
to determine the possible age ranges for the two shell-like features with CVs (Te-11 and AT Cnc).  
In Appendices A-C, details on positions and object types of the three cases studied are presented. 
Appendix D gives details on the nova shell expansion model.
Appendix E provides background on the East Asian records. 
Connections between nova shells and historical guest stars were studied by 
Warner (2016), based on a list of Nickiforov (2010), 
and by Hoffmann (2019), but in an unsatisfactory manner 
(see Neuh\"auser et al. 2021a and Neuh\"auser \& Neuh\"auser 2021).

\section{Nova in M22 and the `guest star' in BC 48}

An extended source (at ca.\,18:36:24.5, $-23$:54:32.4, J2000.0) in the 
globular cluster 
(GC) M22 was presented as nova shell 
and connected to the BC 48 `guest star' based on the supposed rough positional coincidence 
(G\"ottgens et al. 2019).
Its mass, 1-17$\cdot 10^{-5}$\,M$_{\odot}$ (G\"ottgens et al. 2019)
would be normal for nova shells;
however, 
PNe
of multiple stars can also reach down to
$10^{-4}$-$10^{-3}$\,M$_{\odot}$ (Corradi et al. 2015).
The nature or brightness of the central star is unknown,
and so the age cannot even be estimated by a long-term decay rate.
Planetary Nebulae
are rare in GCs, but M22 is one of few GCs with PNe.
It is not yet possible to determine,
whether this new extended feature in M22 is a PN or a young nova shell.
Though there are many 
WDs in GCs, novae in GCs tend to be too distant and faint for the naked eye.
Their shells are too short-lived for detection
because the density of the intracluster medium is too low; 
for example,
in M22, it is only ca.\,0.1 particles cm$^{-3}$ (Monaco et al. 2004).

The Han shu (Tianwen zhi, ch. 26) from China reports for BC 48, fourth lunar month (sometime May 3-31): 
\begin{quotation}
\noindent `First year of Chuyuan reign period under Emperor Yuan [reigned BC 48-32] ... \\
A guest star (ke xing), 
form [or: size] like a melon, 
color blue-white, 
it was 
located at Nandou['s] second star, 
[to the] east, about 4 chi (zai Nandou di er xing dong ke si chi).' 
\end{quotation}
See 
Xu et al. (2000) for the Chinese text;
see Appendix E for background. It was considered a nova in Ho (1966) and references therein.

Based on
all lunar and planetary conjunctions with Nandou stars in the Jin dynasty (AD 265-420, Ho 1966) 
and flanked by further historically based arguments, 
`Nandou['s] 
second star' 
is $\tau$ Sgr, and not $\lambda$ Sgr (Table A.1). 
M22 is outside the newly derived positional area of the `guest star'. 
Since the expected nova peak in M22 has only m=6.4$\pm$1.4 mag (Table 1), 
a
serendipitous naked-eye discovery, color report (`blue-white'), 
and description 
(`form like a melon'), 
i.e. relatively bright and extended, would not be reliable. 
This historical transient does not seem nova-like, 
and in any case
no suitable CV candidate 
has been
found 6$\pm$1.6$^{\circ}$ (`4 chi') east of $\tau$ Sgr (Fig. A.1). It could have been a comet.
For details on 
the
position and object type, see Appendix A.

\section{Te-11 and the `guest star' of AD 483}

The bipolar nebula Te-11 (CRTS J054558.3+022106) 
was interpreted as 
a
nova shell by Miszalski et al. (2016, henceforth M+16),
but it is also still considered 
to be a
PN (e.g. Weidmann et al. 2020).
Its mass (2$\cdot 10^{-4}$\,M$_{\odot}$, M+16) is normal for nova shells, 
but PNe of multiple stars can also have $10^{-4}$-$10^{-3}$\,M$_{\odot}$ (Corradi et al. 2015).
In some cases, there is even a WD inside the PN
(e.g. Wesson et al. 2008, Adam \& Mugrauer 2014). 
The expansion velocity 
of
$\le$10\,km\,s$^{-1}$ of Te-11 (M+16) requires strong deceleration:
it
should be past the Sedov-Taylor phase,
and would then currently dissolve into the interstellar medium (ISM).
Strong deceleration cannot be explained by 
larger-than-normal 
ISM density:
Te-11 is located outside of Orion B and any other clouds (e.g. Lombardi et al. 2014);
its extinction is A$_{\rm V}$=1.0 mag (M+16), which is not unusually large for its Gaia distance
(526$^{+48}_{-47}$ pc, Table 1).

\begin{figure*}
\sidecaption
\includegraphics[angle=0,width=12cm]{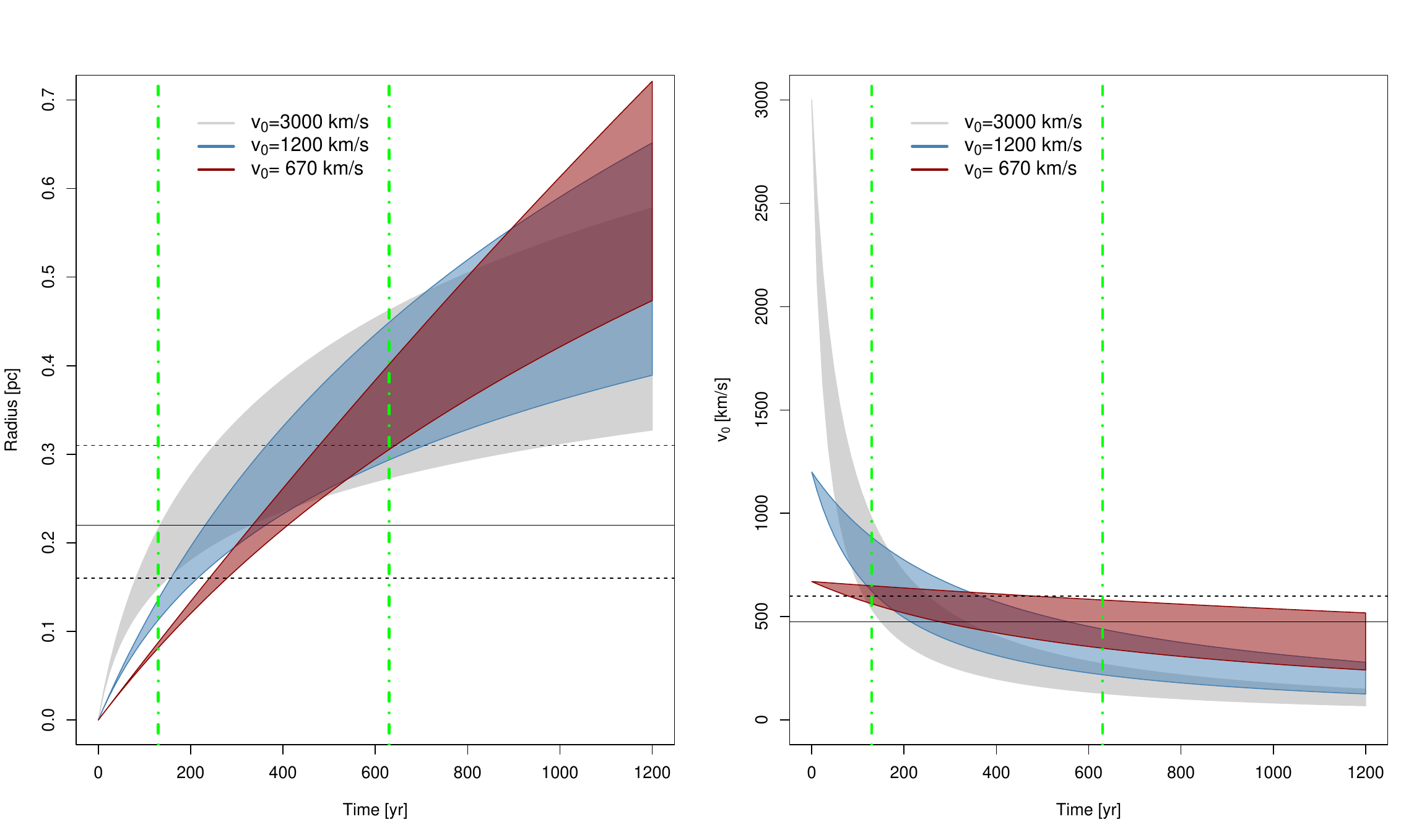}
\caption{Expansion of the AT Cnc ejecta blobs was calculated
following the fluid drag equation (Williams 2013).
For input parameters and details, see Sects. 4 and 5 and Appendix D.
The solution ranges are shown for initial expansion velocities of 
3000 (grey), 670 (red), and 1200 km s$^{-1}$ (blue), 
the latter to differentiate between fast and slow novae.
(We show 670 km s$^{-1}$, the minimum initial expansion velocity
for Sedov-Taylor solutions, for better comparison with Fig. 2.)
Left: Evolution of shell radius.
Right: Evolution of expansion velocity.
The current radius and velocity (in 2016) of the AT Cnc ejecta (horizontal lines) 
are reached simultaneously for an age range of ca.\,128-631 yr 
(dash-dotted green
lines).}
\end{figure*}

\begin{figure*}
\sidecaption
\includegraphics[angle=0,width=12cm]{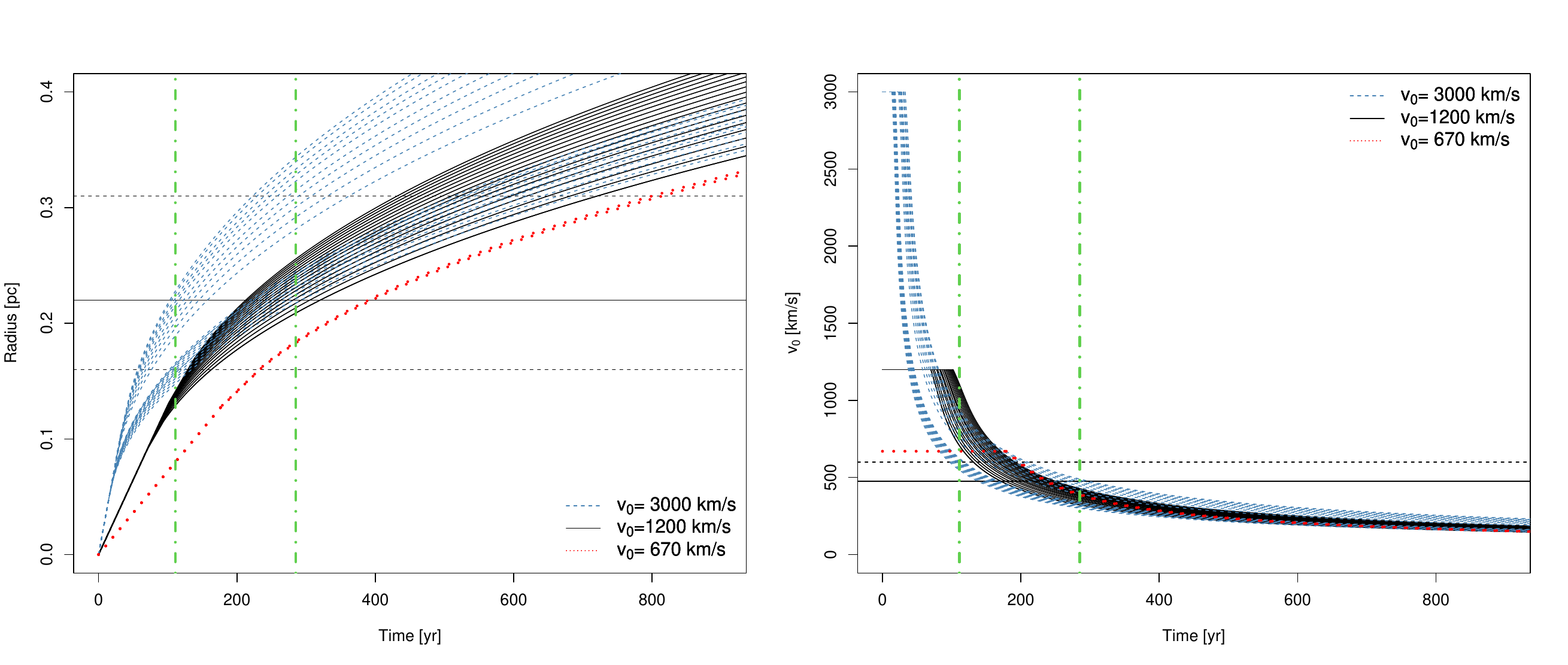}
\caption{Expansion of AT Cnc's nova shell through free expansion and Sedov-Taylor phase with supersonic model calculations:
For input parameters and details, see Sect. 4 and 5 and Appendix D. We use the same initial velocities as in Fig. 1.
Left: Evolution of the shell radius.
Right: Evolution of the expansion velocity.
Both the current radius and the current expansion velocity (horizontal lines) are reached simultaneously
only for an age range of ca.\,111-285 yr 
(dash-dotted green
lines).}
\end{figure*}

The central star is an eclipsing CV with a WD with ca.\,1.18\,M$_{\odot}$ (M+16), 
which is about two times larger than for central stars of PNe.
As a nova it would be fast and possibly recurrent,
but only one historical sighting is suggested despite its prominent location in Orion (Fig. B.1). 
The H$\alpha$ flux is 2-3 orders of magnitude larger than for nova shells older than ca.\,100 yr (Tappert et al. 2020), 
but typical for PNe (see Frew et al. 2016). As 
is
mentioned in M+16, CVs have also been found in PNe.

Since the central star is eclipsing, M+16 could determine 
the
distance (330$\pm$50 pc) and temperature (ca.\,13000$\pm$2000 K)
of the WD, which would then be ca.\,1.5 Gyr old.
However, there are two problems with this scenario:
First, the Gaia distance is much larger (526$^{+48}_{-47}$ pc, Table 1).
Second, Te-11 shows 
HII, 
and therefore requires a photo-ionizing source hotter than ca.\,60000 K (M+16).
This may point to a young WD 
(a few million years old), 
whose progenitor could have produced a PN.
The larger distance cannot solve the temperature problem, 
so a third 
star is still needed (M+16).
There are then no severe problems for a PN interpretation,
while the expansion velocity might be too low for a nova shell.
See also Sect. 5 and Appendix D.

Te-11 with its shell was connected to a guest star in AD 483 based on positional congruency (M+16).
The Wei shu (Tianxiang zhi, ch. 105) from China reports for AD 483, tenth lunar month (sometime Nov 16 to Dec 14): 
\begin{quotation}
\noindent `seventh year of the Taihe reign period under the Emperor Wen [reigned 471-500] of the Northern Wei dynasty [386-534] ... \\
Before this, year 7, month 10, there was a guest star (ke xing), 
form [or: size]
like a dipper (dou), it was 
located 
in/at Shen['s] east (zai Shen dong), similar to 
a radiance\footnote{Liang (1993) gives “radiance” as 
one possible translation for `bo', which concurs with Wang (1999) regarding its basic sense.} (bo).'
\end{quotation}
See Xu et al. (2000) for the Chinese text; see also Appendix E.

The analysis of the brief positional statement reveals that the position could fit, but it remains highly imprecise. 
The complex description of the `guest star' mirrors an irregularly (dipper-like) extended and bright (`bo') transient. 
This and the context in the historical record point to a sighting of a bolide.
As a nova, a peak of only m=2.6$\pm$1.4 mag is expected, so that apparent extension cannot be explained by extreme brightness.
For details on 
the
position and object type, see Appendix B.

\section{AT Cnc and the `large star' of AD 1645}

The ejecta of AT Cnc, a Z-Cam-type 
dwarf nova, 
were studied by Shara et al. (2012b, henceforth S+12b; 2017a, henceforth S+17a): 
the 
mass of $\le$\,5$\cdot 10^{-5}$\,M$_{\odot}$ is near the lower range for nova shells;
the radius is r=$0.22^{+0.09}_{-0.06}$ pc at 460 pc, confirmed by Gaia.
The current radial velocities of the ejecta blobs show a maximum expansion velocity of v=475 km s$^{-1}$;
given the inclination, the radial velocity may be up to $\sim$550-600 km s$^{-1}$.
S+17a: `If we ignore deceleration of the ejecta, we find T=489$^{+200}_{-133}$yr' as age
(for epoch 2016).

Downes \& Duerbeck (2000) and Santamaria et al. (2020) 
analysed six distinct nova shells (see Sect. 1 under point 3)
and showed that they do not decelerate during the free expansion. 
However, S+17a assumed that the AT Cnc shell would have passed through two 
velocity-halving timescales
of ca.\,75 yr each (Duerbeck 1987).
Then, its initial expansion velocity would have been
$\le 550 \cdot 2 \cdot 2 = 2200$ km s$^{-1}$ some $330^{+135}_{-90}$ yr ago,
i.e. during AD 1551-1776.

To estimate
the age of the Sedov-Taylor phase with $t=0.4 \cdot r/v$, 
the initial velocity (in 
the
case of no deceleration during the free expansion) and the radius 
at the start of the Sedov-Taylor phase must be known.
If the initial velocity was 2200 km s$^{-1}$ and if the free expansion lasted 
a few decades (S+17a), then
the radius at the end of the free expansion 
would be
$\sim$0.11 pc. 
Then, the Sedov-Taylor phase would be shorter than 
what is
given above.

In Sect. 5, a full calculation of nova shell evolution by two methods with all relevant parameters 
(including ISM density) is performed to re-estimate the age range.
For AT Cnc, a shell age of 128-631 yr (fluid drag of ejecta blobs, Fig. 1)
or 111-285 yr (free expansion plus Sedov-Taylor phase, Fig. 2) is obtained.
For details on these models, see Sect. 5 and Appendix D.

Following Warner (2016), based on Nickiforov (2010),
a link to a Korean record was considered due to the estimated shell age and positional coincidence (S+17a).
The Chungbo munhon pigo (ch. 6) from Korea reports for AD 1645, 
second month (sometime Feb 26 to Mar 27): 

\smallskip

\indent `Twenty-third year of the Korean sovereign Injo [reigned \\
\indent 1623-1649] ... \\
\indent A large star (da xing) entered (ru) Yugui.' 

\smallskip

\noindent See Xu et al. (2000) for the Classical Chinese text and Appendix E for details.

The position of AT Cnc is 
not closely related to the lunar mansion asterism Yugui (ca.\,$5^{\circ}$ separation), 
and it does not lie in the much larger RA range of lunar mansion Yugui,  
but rather in the RA range of lunar mansion Dongjing (Fig. C.1). 
The verb (`ru'=`enter') implies motion (Ho 1966, p. 37):
entering a well-defined field
is often recorded in the context 
of
lunar and planetary movement.
An examination of the record reveals that the `large star' (not `guest star')
might have been the planet Mars, which `entered' the 
four-star polygon Yugui.
There are several cases around that time 
in which
the same Korean source 
also did not identify Venus using its conventional name, 
but reported it as a `monstrous star'.
For details on 
the
position and object type, see Appendix C.

\section{Nova shell expansion of AT Cnc}

The expansion age of a nova shell 
has to agree with the time since the suggested guest star observation. 
The duration of the expansion required
to reach the current radius and expansion velocity
depends on the following:
the
ejected mass, ISM density, and initial expansion velocity ($\sim$300-1200 km s$^{-1}$ for slow 
and $\sim$1200-3000 km s$^{-1}$ for fast novae, Yaron et al. 2005).
We have performed a large number of nova shell expansion calculations 
for AT Cnc (Figs. 1 \& 2) and Te-11 (Fig. D-1 in Appendix D).

One can estimate the expansion age either by the fluid drag equation, 
if ejecta blobs are observed (Williams 2013, Santamaria et al. 2019), 
or by considering the shell evolution with free expansion,
a Sedov-Taylor phase, and a snowplough 
phase (Taylor 1950, Ostriker \& McKee 1988, Shu 1992, Truelove \& McKee 1999, Padmanabhan 2001); 
see Appendix D for details.

If nova shells behave 
similarly to 
scaled-down supernova remnants with 
a
reduced ejected mass and velocity, and hence with shorter lifetimes,
the nova shell accelerates to speeds greater than the speed of sound (supersonic). This causes
the shock wave to move in front of the ejecta. The high-velocity material ploughs outwards into the ISM, 
compressing and heating ambient gas and sweeping it up 
similarly to a snowplough. 
The blast collects the expelled material
and any additional material from the ISM 
as it travels through it, thereby forming a nova shell.
Free expansion ends roughly when the ejected mass equals the swept-up mass, depending on 
the
ISM density and initial expansion velocity.
According to Santamaria et al. (2020), deceleration during the free expansion phase is negligible.

For AT Cnc, the Gaia DR3 geophotocentric distance 456$\pm$6 pc (Bailer-Jones et al. 2021) is used,
together with the current radius of the shell-like structure of $0.22^{+0.09}_{-0.06}$ pc (S+12b),
the current expansion velocity of 475-600 km s$^{-1}$ (S+17a), 
an initial ejection velocity of 475-3000 km s$^{-1}$ (1 km s$^{-1}$ steps),
and 
a
normal ISM density of 0.1 to 1 cm$^{-3}$.
For the ejected mass, 21 values from 1.0 to 5.0$\cdot 10^{-5}$\,M$_{\odot}$ were used
(S+12b obtained the latter as 
an
upper limit for the total shell mass, ejecta plus swept-up ISM;
the lower bound is the ejected mass from Yaron et al. 2005 for the parameters of AT Cnc).

In the case of AT Cnc, 
in which
115 discrete ejecta blobs (S+12b) move through the ISM,
their motion can be described by the drag equation (Williams 2013, Santamaria et al. 2019): 
the evolution of the radius, $r$, velocity, $v$, and acceleration, $a$, 
of the ejecta shell with time $t$ depend on the physical parameters of 
the ejecta blobs (given in S+12b) and the ISM density and follow $a \propto -v^{2}$, $r \propto log(t)$, and $v \propto 1/t$.
A large number of simulations were
performed for the initial velocity of the blobs in the range of 475 to 3000 km s$^{-1}$ 
and ISM density of 0.1 to 1 cm$^{-3}$. 
The ejecta can then reach the current radius (0.16-0.31 pc) and velocity (475-600 km s$^{-1}$, S+17a)
simultaneously at an age range of ca.\,128-631 yr (Fig. 1).

If the expansion age of the supersonic shock wave is estimated with free expansion and 
a
Sedov-Taylor phase,
the combined age (free expansion plus Sedov-Taylor) is 52-800 yr 
to reach the current radius (0.16-0.31 pc);
and 97-305 yr would be allowed to reach the current velocity range of 475-600 km s$^{-1}$.
Since both conditions have to be fulfilled simultaneously, the 
age range is constrained to 111-285 yr (Fig. 2) 
and the initial expansion velocity to $\ge 670$ km s$^{-1}$.

The free expansion without deceleration here lasts ca.\,40-60 yr; 
the duration of the Sedov-Taylor phase alone is then consistent 
with the approximation by $0.4 \cdot r/v=181^{+74}_{-57}$ yr (S+17a).
For an initial ejection velocity 
of
$\le$1200 km s$^{-1}$, i.e. a slow nova, 
an age range of ca.\,149-250 yr is obtained to reach 
the
current radius and velocity simultaneously.
For a fast nova, the whole range of ca.\,111-285 yr is possible (for certain parameter combinations).
A transient in AD 1645 would be too early here,
unless the ejected mass is <10$^{-5}$\,M$_{\odot}$ or the ISM density is >1\,cm$^{-3}$.
However, the extinction towards AT Cnc is very small
(see Table 1).

The duration of visibility for a drop of 0.13$\pm$0.05 mag day$^{-1}$ for fast novae (Della Valle \& Izzo 2020, therein table 2)
would have been 36$\pm$11 days (from a peak of m=1.3$\pm$1.4 mag to 6.0 mag, the naked-eye limit).
A slow nova would have lasted up to ca.\,$8 \pm 3$ months (for a drop of 0.013-0.024 mag day$^{-1}$, Della Valle \& Izzo 2020).
So far, the earliest classical novae
brighter than 3.0 mag, and observed also telescopically, were V841 Oph in 1848 (no shell), T CrB in 1866, 
and T Aur in 1891. In principle, a reported visibility duration or a WD mass (too imprecise 
in the case of AT Cnc, see Table 1) 
can constrain the nova speed class or vice versa.

Extrapolating the empirical mean decay rate of 10$\pm$3 mmag yr$^{-1}$ for novae from the last century (Duerbeck 1992), 
which is consistent with expectation from hibernation (Kovetz et al. 1988), 
would suggest that 
AT Cnc (expected peak at m=1.3$\pm$1.4 mag, now at 12.0-16.6 mag, Table 1) 
had exploded 1070$\pm$284 to 1530$\pm$380 yr  
before the 2016 epoch of S+17a, namely, ca.\,AD 106-1230.
Since this is inconsistent with the derived shell age ranges (Figs. 1 \& 2), 
we can conclude that such a large extrapolation may not be allowed -- 
the decay was much faster.

Given the estimated peak brightness of the AT Cnc nova eruption (m=1.3$\pm$1.4 mag, Table 1) and its prominent 
position in a relatively dark area, a historical observation of a nova 
could be expected 
to have taken place roughly between the 15th and 19th centuries.

\begin{acknowledgements}
DLN and RN designed this study, wrote together the main part, DLN also most of Appendices A-C;
VVH contributed Sect. 5 and Appendix D with RN; JC supported the sinological considerations;
MDV concentrated on the astrophysical nova issues.
All authors read and commented on the full paper.
VVH and RN would like to thank the Deutsche Forschungsgemeinschaft for financial support in grant numbers NE 515/61-1 and 65-1.
VVH acknowledges support by the Yerevan State University in the framework of an internal grant.
We acknowledge Jonna Schickhoff for double-checking the data in Table A.1, 
Veronika Schaffenroth (TLS) for discussion on Te-11,
Matthias Steffen (AIP) for advise on PNe, Ernst Paunzen (U Brno) for help with M22, and
G\"unay Payli and Baha Dincel for discussion on shells.
\end{acknowledgements}

{}

\begin{appendix}

\nolinenumbers

\section{Details on the BC 48 record}

(a) Position: The star chain of the 
lunar mansion asterism 
Nandou (lit. southern dipper) connects $\mu, \lambda, \phi, \sigma, \tau$, and $\zeta$ Sgr 
from handle to bowl/head/box (Fig. A.1). According to Stephenson \& Green (2009), cited by G\"ottgens et al. (2019), 
`available charts' would show that `the second star of Nandou was $\lambda$ Sgr' with reference to Yi (1984) and Pan (1989). 
The modern nomenclature has for Nandou's chain 3, 2, 1, 4, 5, 6 (from $\mu$ to $\zeta$ Sgr), 
so that $\lambda$ Sgr would be star “2” (Fig. A.1) -- only this numbering is available in Yi (1984). 
Ho (1966) relied on this modern use when numbering Nandou's stars 
given in the context with lunar and planetary conjunctions during the Jin dynasty (see below), 
but he overlooked that none of them is then in accordance with the reported observations (see Table A.1). 

There is ample evidence that numbering sequences as implemented in late imperial works (18th century), see Yun (1845, 1983),
which form the basis of modern reference materials on traditional star names
(e.g., the star chart that appears in Wang 1999),
were not in use as of BC 48; Stephenson \& Green (2009): 
`possibly the numbering system adopted during the Han dynasty [BC 206 to AD 220] differed from that in later centuries'.    

The systematic change from pre-modern to modern is well documented in Pan (1989). 
For instance, the undisputed determinative star of 
lunar mansion asterism 
Nandou, namely $\phi$ Sgr, 
is defined in the two systems by clearly distinguishable wording: \\
(1) relying on cardinal numbers the modern system has \mbox{`[Nan-]}Dou xiu 
[i.e. lunar mansion asterism] one' (e.g. 
Yi 1984 and Pan 1989, p. 346); \\
(2) the pre-modern always gives ordinal numbers (if counting at all) together with further descriptive specifications, 
usually `[Nan-]Dou xiu ... [from the] bowl/head/box, [the] fourth star' (e.g., Pan 1989, p. 293) 
or as relatively rare alternative `[from the] west [the] 3rd star' (Pan 1989, p. 214) -- 
here, `[from the] west' is equivalent to `[from the] handle'; see Fig. A.1. \\ 
In order to explicitly mark ordinal numbers in Classical Chinese, one needs an additional character, `di', 
which is followed always by the respective number and then the character for star, `xing'.

The text from the Han shu for BC 48 gives just `Nandou['s] second star' -- 
with ordinal number, but without specification from which end to count.

First, the transmissions that came down to us are late compilations, often shortened and therein also concatenated, 
i.e. characters or whole phrases may be omitted. 
Since some observational records (in addition to the star charts, see Pan 1989) 
still preserve a more original design (e.g., Comet Halley for AD 837, Pankenier et al. 2008), 
the pre-modern notation can be reconstructed.
The transmitted texts show that the pre-modern system was applied by and large in the same manner from (at latest) 
the Han dynasty onward (Sun \& Kistemaker 1997), completed during the Jin dynasty (Jin shu, Ho 1966),
and continuously performed at least into the 18th century (e.g., see Pankenier et al. 2008); 
in principle, this system was also adopted by Korean (see Sect. 4, AD 1645) and Japanese court astronomers.

If the text for BC 48 would be shortened, 
i.e. omitting the specifications `bowl/box/head' or `handle' (or from east or west), 
the `second 
star' could be $\tau$ or $\lambda$ Sgr.

However, already Stephenson \& Green (2009) argued that `allowing for precession, a location approximately $4^{\circ}$ 
[literal: 4 chi] to the east of this 
[second] 
star [$\lambda$ Sgr is assumed here from the modern notation] 
would be only about $2.5^{\circ}$ from $\phi$ Sgr 
(towards the north)', so that the latter should have been specified.
The error circle in Hoffmann (2019) does not fulfill this condition and even encompasses several Nandou stars.

\begin{figure}
\includegraphics[angle=270,width=0.5\textwidth,trim=0 55 0 0]{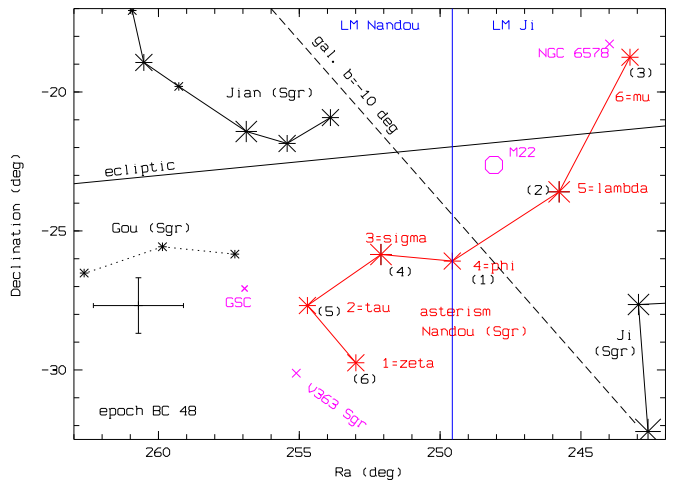}
\caption{The Chinese 
lunar mansion asterism 
Nandou (Sgr); star sizes correspond to brightness;
stars are numbered both according to the new reconstruction found here (red, historical counting,
Table A.1) and the previous assumption (black, modern counting).
The new search area for the BC 48 `guest star' east of Nandou's 
second 
star is shown with uncertainty.
Previously suggested nova counterparts (NGC 6578, V363 Sgr, M22) and also  
the CV GSC 06883-00545 (GSC) are outside this field, see text --
and do not reflect the record of BC 48, which pertains most certainly to a comet sighting.
For the 
two-star 
asterism Guo, the three suggested stars are shown.}
\end{figure}

Are there further hints to the practice of counting in Nandou? 
Could the short text also be complete in the sense that the position as given is nevertheless well-defined? 

Liu (1986) obtained the numbering of stars in Nandou and other asterisms near the ecliptic from 144 lunar 
occultations AD 342-565 (114 as given in the English-language abstract) 
and 76 other lunar conjunctions AD 479-494 as recorded in dynasty chronicles, but this was apparently overlooked by most others. 
For Nandou, Liu (1986) obtained a numbering beginning with $\zeta$ Sgr (i.e. from the bowl) as `first star' 
and then with a continuous numbering including $\lambda$ Sgr as `fifth star'. 
Hoffmann (2019, figure 4) did show a partial scan from Liu (1986), but neglected the numbering given 
for Nandou when discussing the case of M22, so that the claim that M22 `indeed fits roughly the position' 
(presumably close to $\lambda$ Sgr as argued in G\"ottgens et al. 2019) is unsubstantiated. 
Yau (1988) gives $\tau$ Sgr as `second star' without reference, probably based on Liu (1986),
because Liu (1986) is otherwise cited by Yau (1988); 
also Stephenson (1994, p. 530) mentioned the Liu (1986) investigation -- yet, Stephenson \& Green (2009) do not recruit on it.

\begin{table*}
\begin{tabular}{lllcc|llr|lrl} \\ \hline
\multicolumn{11}{l}{Table A.1: Conjunctions of Nandou stars with moon or planet in the Jin shu for the Jin dynasty\tablefootmark{(a)}:} \\ \hline
\multicolumn{3}{l}{Record in Jin shu (Ho 1966)} & \hspace{-0.6cm} Nandou & Ho & \multicolumn{3}{l}{True closest conjunction} & \multicolumn{3}{l}{when visible\tablefootmark{(b)}} \\
Date of night   & Object & Jin shu text             & Star & ID        & Star          & Date LT       & Sep. & Time & Sep. & Note \\ \hline
347 Feb 19[/20] & Moon   & `trespasses against'     & 5th  & $\tau$    & $\lambda$ Sgr & Feb 20, 03:49 & 28'  & same & & \\
347 Jun 9[/10]  & Moon   & `trespasses against'     & 4th  & $\sigma$  & $\phi$ Sgr    & Jun 9, 22:44  & 33'  & same & & (c)  \\
347 Oct 25[/26] & Venus  & `trespasses against'     & 5th  & $\tau$    & $\lambda$ Sgr & Oct 25, 20:10 & 22'  & Oct 25, 19:00 & 23' & (d) \\
348 Sep 16[/17] & Venus  & `trespasses against'     & 2nd  & $\lambda$ & $\tau$ Sgr    & Sep 16, 15:49 & 30'  & Sep 16, 19:30 & 82' & (e) \\
352 Apr 24[/25] & Moon   & `trespasses against'     & 2nd  & $\lambda$ & $\tau$ Sgr    & Apr 21, 00:47 & occ. & same & & \\ 
352 Oct 24[/25] & Venus  & `trespasses against'     & 4th  & $\sigma$  & $\phi$ Sgr    & Oct 24, 10:28 & 22'  & Oct 24, 18:40 & 27' & \\ 
353 Apr 11[/12] & Moon   & `trespasses against'     & 3rd  & $\mu$     & $\sigma$ Sgr  & Apr 10, 21:59 & occ. & Apr 10, 23:30 & 76' & (f) \\ 
360 Oct 8[/9]   & Venus  & `trespasses against'     & 4th  & $\sigma$  & $\phi$ Sgr    & Oct 25, 04:52 & 26'  & Oct 24, 18:42 & 33' & (g) \\
373 Apr 27[/28] & Moon   & `concealed'\tablefootmark{(k)} & 5th  & $\tau$    & $\lambda$ Sgr & Apr 26, 21:31 & 16'  & Apr 26, 22:00 & 22' & \\
376 May 21[/22] & Mars   & `trespasses against'     & 3rd  & $\mu$     & $\sigma$ Sgr  & May 22, 23:55 & 11'  & same & & \\ 
376 May 31[/J1] & Mars   & `concealed'\tablefootmark{(k)} & 4th  & $\sigma$  & $\phi$ Sgr    & Jun 1, 03:12  & 3'   & same & & \\
403 Jul 17[/18] & Moon   & `concealed'\tablefootmark{(k)} & 4th  & $\sigma$  & $\phi$ Sgr    & Jul 17, 20:40 & occ. & same & & \\
404 Jun 9[/10]  & Moon   & `concealed'\tablefootmark{(k)} & 2nd  & $\lambda$ & $\tau$ Sgr    & Jun 10, 03:26 & occ. & same & & (h) \\
405 Sep 17[/18] & Moon   & `trespasses against'     & 1st  & $\phi$    & $\zeta$ Sgr   & Sep 17, 17:10 & 57'  & Sep 17, 19:20 & 74' & \\
406 Jul 27[/28] & Mars   & `trespasses against'     & 5th  & $\tau$    & $\lambda$ Sgr & Sep 18, 22:26 & 19'  & Sep 18, 21:25 & 19' &(i)\\
408 Jun 23[/24] & Moon   & `concealed'\tablefootmark{(k)} & 2nd  & $\lambda$ & $\tau$ Sgr    & Jun 23, 20:17 & 34'  & same & & (j) \\
410 May 6[/7]   & Moon   & `concealed'\tablefootmark{(k)} & 5th  & $\tau$    & $\lambda$ Sgr & May 7, 04:41  & occ. & same & & (h) \\ 
410 Jun 30[/J1] & Moon   & `concealed'\tablefootmark{(k)} & 5th  & $\tau$    & $\lambda$ Sgr & Jun 30, 23:48 & 17'  & same & & \\  
410 Sep 20[/21] & Moon   & `trespasses against'     & 5th  & $\tau$    & $\lambda$ Sgr & Sep 20, 17:58 & occ. & Sep 20, 19:20 & 36' & \\ \hline
\end{tabular}
\tablefoot{
\tablefoottext{a}{For 19 lunar and planetary conjunctions and occultations (occ.) with Nandou stars (1st, 2nd, ...) recorded in the Jin shu 
for the Jin dynasty (AD 265-420), cited after the English translation with commentary by Ho (1966), 
the star identification (ID) within Nandou (in Sgr) as given therein
(following modern counting of star numbers) is listed,
then the stellar identification of the historical observation for the closest conjunction as determined with Stellarium version 0.18.1 
and confirmed with JPL ephemeries (using $\Delta$T from Morrison \& Stephenson 2004); 
dates and times given in Local Time (LT) 
for the Jin dynasty capital Jiankang, now Nanjing, China, 7:55h east of UT
(Chinese records give the date of the evening for the whole night). 
Always, the closest conjunction with a Nandou star in the recorded night is considered, 
or otherwise the closest conjunction with a Nandou star within a few nights (Sep. for separation). 
In sum, this study of the reported conjunctions yields a fully consistent numbering of the Nandou stars,
which can therefore be accepted as historical counting: 1st -- $\zeta$, 2nd -- $\tau$, 3rd -- $\sigma$, 4th -- $\phi$, 5th -- $\lambda$ Sgr.
However, the identifications in Ho (1966) never follow the observations.} \\ 
\tablefoottext{b}{The following conditions are employed: The observed conjunctions have to be at least $5^{\circ}$ above horizon at darkness,
i.e. when the Sun is at least $18^{\circ}$ below horizon. If the closest conjunction is at day-time or below horizon,
we list the next closest that fulfills the conditions for observability.} \\
\tablefoottext{c}{For completeness, there was also an occultation of $\sigma$ Sgr on June 10 at 4:23h shortly after moon-set.} \\
\tablefoottext{d}{Closest conjunction was less than one hour after Venus had set; a close conjunction as reported was visible the hours before.} \\
\tablefoottext{e}{On that day, it was the Moon, not Venus, trespassing against $\tau$ Sgr.} \\
\tablefoottext{f}{Also a close conjunction with $\tau$ Sgr on Apr 11 visible at darkness above horizon at 4:22h with 77' separation.} \\
\tablefoottext{g}{Here, the date in Jin shu is off by some 2.5 weeks.} \\
\tablefoottext{h}{On 404 June 9, the Sun was at $-15^{\circ}$, on 410 May 6 at $-6^{\circ}$ altitude, 
both in the NE during the closest lunar conjunction in the SW.} \\
\tablefoottext{i}{Here, two observations seem to have been mistakenly merged into one: On the given date, Jul 27, Mars was in conjunction with $\delta$ Sco,
while a close conjunction of Mars with $\lambda$ Sgr was on Sep 18 of the same year; no conjunction with $\tau$ Sgr that year.}  \\
\tablefoottext{j}{During the close conjunction of $\tau$ Sgr on June 23 at 20:17h, the Sun was $13^{\circ}$ below horizon in the W;
during daytime at 11:42h (below horizon), there was an occultation of $\phi$ Sgr by the moon.} \\
\tablefoottext{k}{Of the seven conjunctions described with `concealed' (AD 265-420),
six were very close encounters (or even occultations) between Moon and a Nandou star,
one was the closest planetary conjunction (Mars, 3$^{\prime}$).} 
}
\end{table*}

The work of Liu (1986) is updated here with all planetary and lunar conjunctions with Nandou stars 
recorded by the Jin dynasty (Ho 1966); it is shown that a continuous numbering is perfectly performed (Table A.1): 
The counting in Nandou started with $\zeta$ Sgr as 
the first star, then $\tau$ Sgr as the second star, 
then further along the skeleton line until (at least) $\lambda$ Sgr as 5th star 
($\mu$ Sgr is the only one for which no such observations are given in the Jin shu), 
i.e., first the `bowl/head/box', then the `handle' (Fig. A.1). 
This numbering structure is similar in Nandou's northern counterpart Beidou (`northern dipper', 
identical with the western Big Dipper) as explicitly given in Jin shu (Ho 1966). 
The Nandou counting obtained for the Jin dynasty observations AD 347-410 quite certainly also holds for the Han dynasty, 
i.e. during the observation of the BC 48 guest star,
when this system was basically fixed (Sun \& Kistemaker 1997). 
Counting in such a continuous sequence was performed until at least the 13th century, 
also, 
e.g. in 
Korea (Stephenson 1994, p. 559, see also Koryo sa, ch. 49): 
In AD 1223 [Nov 4] `Venus invaded the fifth star of Nandou', 
Nandou's fifth star is clearly $\lambda$ Sgr (separation $\le 1^{\circ}$). 

(b) Object type: 
Neither the size/form nor the color given for the transient of BC 48 supports an appearance like a classical nova: 
`form [or: size] like a melon' 
(whatever it exactly means) seems to reflect on a significantly extended angular size 
and/or maybe large brightness. 

For a typical nova absolute brightness of $-7.0 \pm 1.4$ mag (Schaefer 2018, Della Valle \& Izzo 2020) 
and for the distance of M22 (3$\pm$0.3 kpc, Monaco et al. 2004)
and extinction to M22 (Table 1), the expected nova peak brightness would be m=6.4$\pm$1.4 mag -- 
far too faint to be discovered serendipitously by the naked eye.

The color of the `guest star' was reported as `blue-white'. 
Novae at maximum have an intrinsic color index of (B-V)$_{0}$=0.23$\pm$0.06 mag with a dispersion of 0.16 mag 
(van den Bergh \& Younger 1987, similar in Seitter 1990), 
or (B-V)$_{0}$=0.11$\pm$0.04 mag, as given by Schaefer (2023) based on a much larger sample (504 Galactic novae) including more recent CCD data.

If the object were a CV, one could assume (B-V)$_{0} \approx 0.0$ mag for quiescence. 
For its extinction of A$_{\rm V}$=1.0 mag (references in Table 1), the normal reddening law of R=3.1, 
and typical color indices of novae at peak, one can expect an apparent color index of B-V$\approx$0.43-0.55 mag
(i.e. within the range for `green', B$-$V=0.3-0.6 mag). 
This would appear color-less whitish to the naked eye, 
but the report has `blue-white' color.
A `blue' color was reported in the Han dynasty, e.g., correctly for Bellatrix (B-V=$-0.14$ mag), 
and a `white' color, e.g., for Sirius (B-V=$0.01$ mag), see Neuh\"auser et al. (2022).
A `blue-white' color was often reported for comets pointing to a white head (plus possibly a whitish dust tail) and a blue plasma tail.

Although the hour angle between Sun and Nandou (Sgr) is about 7h in May, 
some comet criteria are fulfilled (see D.L. Neuh\"auser et al. 2021b, therein section 2.4): 
position close to the ecliptic (latitude ca.\,$5^{\circ}$), color `blue-white' (plasma tail seen) 
and similar description of extension 
(`form [or: size] like a melon')
can otherwise be found in credible comet reports (e.g. BC 147 Aug 6 
`form [or: size] like a peach', our translation; for Chinese text, see Pankenier et al. 2008). 
During the Han dynasty, several obvious comets were given as 
guest stars, 
e.g. BC 49, BC 47, AD 55, AD 66 (Pankenier et al. 2008). 
Unfortunately, neither the exact date of detection (only the lunar month) nor duration or possible motion 
of the BC 48 transient are transmitted. 
The above mentioned counterargument regarding a comet interpretation, an hour angle of 7h, may not be valid:
In compilation processes, it can happen that the given date could reflect the first appearance, 
while the remaining text belongs to later phases of the event or vice versa, 
see Neuh\"auser et al. (2021a) for the comet of AD 891.

In principle, the measurement precision to the exact `chi' may not be credible for a short-term 
and/or rapidly moving object like a bright meteor or bolide, but it speaks for a more static phenomenon,
e.g. a comet or a mock moon; 
a left or right paraselene would lie by and large in the given positional area
in the nights BC 48 May 17/18 or 20/21 (sufficiently close in time to full moon). 

One could take into consideration religious and cosmological criteria: 
In ancient China, specific celestial phenomena lead to well-documented different astro-omenological interpretations -- 
but caution is needed (not only) because the patterns are based on pheno-typical descriptions and do not mirror
automatically the physical nature of the object; 
yet, the bias of a retroactive reading in the sense of self-fulfilling prophecy might generally not be substantiated. 
Continuing the observational report, the source comments on its omenological interpretation: \\
`The prognostication says: “Famine on account of flooding.” In the fifth month, the waters of Bohai greatly burbled up. 
In the sixth month, there was a great famine east of the pass [Guandong]. Many people starved to death. 
In Langye Commandery, people ate one another.'

The transient of BC 48 is omenologically associated with a famine due to a flood.  
The occurrence of the flood and the famine may have influenced early historians' sense of the significance (and magnitude) 
of the astronomical event. The Kaiyuan zhanjing (79.3a), a high Tang source (compiled during the Kaiyuan reign period, AD 713-742) 
known to preserve essential information from earlier sources, 
has only: `In month 5, the waters of Bohai greatly overflowed.' 
Both auspicious and inauspicious associations with comet sighting are well-established and not limited to 
East Asian traditions (Loewe 1994, Raphals 2013, Chapman 2021); 
on the other hand, halo phenomena like the 22$^{\circ}$-ring are meteorologically indeed a precursor for rainfall --
however, a sole paraselene hardly indicates excessive flooding.

(c) Final remarks: 
Even though the analysis of the object type (see b) refers
clearly to a non-stellar transient, we will discuss briefly (i) the position of the error box,
(ii) previous counterpart suggestions, and (iii) the `guest star' of AD 1011 nearby.

(i) The BC 48 guest star in Nandou was observed `about 4 chi' east of $\tau$ Sgr. 
Since one chi corresponds to 1.50$\pm$0.24$^{\circ}$ (Kiang 1972a,b) 
and since the separation was specified to (a precision of) one chi only, given as `about 4 chi' (say 4.0$\pm$0.5 chi),
an additional uncertainty of $\pm 0.5$ chi was added in quadrature, resulting in 6$\pm$1.6$^{\circ}$ east of $\tau$ Sgr. 
The declination can be further constrained to be closer to $\tau$ Sgr than to any other Nandou star (Fig. A.1). 

NB: The Nandou asterism, as one of the 28 
lunar mansion asterisms, 
was of high relevance.
The error box is also situated south of the faint two-star asterism Gou (Fig. A.1).
It is not from the main Shi Shi, but the minor Gan Shi school --
its position is described in relation to the Nandou asterism (Sun \& Kistemaker 1997, p. 180);
Guo is not mentioned in relation to guest stars and comets (see indices in Xu et al. 2000 and Pankenier et al. 2008).
The identification of Gou's two stars is uncertain: Yi (1984) and Ho (1966) gave $\chi^{1}$ Sgr and h2/h1 Sgr, while
Sun \& Kistemaker (1997) plotted $\chi^{1}$ Sgr and $\psi$ Sgr (charts in the back of their book, without page numbers), 
but gave 51 Sgr (h1 Sgr) as its main star (appendix, p. 181). All are fainter than 4.5 mag. 
All three suggestions are plotted in Fig. A.1.

(ii) Previously suggested possible counterparts to the BC 48 guest star as nova/SN are all outside the error box (Fig. A.1):
PN NGC 6578 (Hsi 1957), supernova remnant (SNR) G21.5-0.9 (Wang et al. 1986, 2006, outside Fig. A.1),
V363 Sgr (Yau 1988, but with its Gaia EDR2 geometric distance of $5.0 \pm 2.6$ kpc (Bailer-Jones et al. 2021) 
it is too distant for a naked-eye sighting, its brightest magnitude in the 1927 eruption was 8.8 mag), 
and M22 (G\"ottgens et al. 2019, discussed here, see Sect. 2 and Table 1). 

For completeness,
it was checked whether any nearby CVs, novae, dwarf-novae, nova-likes, symbiotics, WDs, including candidates as well as pulsars and SNRs 
are in the BC 48 error box in Simbad, but none were found.

According to the Simbad database, there is only one nova counterpart candidate near (but outside) the revised BC 48 error box,
which fulfills the criteria from Table 1 (naked-eye detectability as nova, expected nova peak and amplitude): \\
GSC 06883-00545 has a Gaia DR3 parallax of $1.32 \pm 0.044$ mas (RUWE 1.118) with a photogeometric distance of
$730 \pm 23$ pc (Bailer-Jones et al. 2021); for an extinction of A$_{\rm V}=0.4$ mag (Yu et al. 2023),
we expect a nova peak from M=$-7.0 \pm 1.4$ mag to be $2.7 \pm 1.4$ mag;
and with the current G=15.6 mag, we expect an amplitude of ca.\,12.9 mag; 
hence, a naked-eye sighting of GSC 06883-00545 as Classical Nova would just be possible. 
This object is shown in Fig. A.1.
(For another one with sufficient peak brightness, DENIS J191810.6-262106, a WD without known binarity, the amplitude as nova would be too large.)
There are no further nova candidates, which would be detectable by the naked eye, i.e. being within ca.\,2 kpc 
(the limit for serendipitous discovery for a peak fainter than 3 mag with M=$-7.0 \pm 1.4$ mag for novae, for negligible extinction).

We stress that the `guest star' of BC 48 was not of stellar nature (probably a comet), so that none of the above can be a counterpart.

(iii) There is another guest star observation, AD 1011, whose position is broadly consistent with both
the BC 48 error box and GSC 06883-00545: \\
AD 1011 Feb 8: `A guest star appeared in front of the bowl of Nandou' \\
(Xu et al. 2000, pp. 138 and 333, from Song shi).
Since Guo is also described that way (Sun \& Kistemaker 1997, p. 181) and always depicted east of Nandou,
the guest star of AD 1011 is somewhat close to the Guo stars.
Hence, GSC 06883-00545 would a viable candidate.
The AD 1011 record has `guest star' and also gives a precise date,
but the position is close to the ecliptic and, at the given date, even close to the Sun --
and therefore more probably points to a comet sighting.
There is no information on duration, form, size, or color, so that caution is needed.

\section{Details on the AD 483 record}

(a) Position: 
The seven main stars of the 
lunar mansion asterism 
Shen are the same as of the basic figure of Orion 
($\alpha, \beta, \gamma, \kappa, \delta, \epsilon, \zeta$ Ori). 
Te-11 lies slightly west of the skeleton line from $\alpha$ to $\zeta$ Ori and nearer to the latter (Fig. B.1). 
Miszalski et al. (2016) assumed a positional congruency from the observational text as given in Xu et al. (2000). 
It was suggested that the position of the transient in AD 483 would be situated in a huge area 
east 
of the eastern asterism lines (Stephenson \& Green 2009), 
possibly within the right ascension (RA) range of 
the lunar mansion
Shen (e.g. as explicitly in Hoffmann 2019, figure 9).\footnote{The 
lunar mansion RA range 
Shen (LM Shen) runs from the
determinative star $\delta$ Ori at $\alpha$=4h 15m 25.0s
($\zeta$ Ori is discussed only for the 17th century, see Stephenson 1994) to
the determinative star of the next LM (Dongjing) being $\mu$ Gem at $\alpha$=4h 51m 43.0s; 
Te-11 had $\alpha$=4h 27m 43.2s, all epoch AD 483.}
Then, the derived area would not include Te-11.

However, the close examination of the brief positional statement `zai Shen dong' offers the possibility of coincidence.

In the context of observational transmissions, 
the otherwise very common verb `zai' introduces quite a precise position
(e.g. D.L. Neuh\"auser et al. 2021b): \\
If `zai' comes with a `du' measurement (1 du is about 1 degree) from a specific determinative star, 
it gives an hour angle and, hence, a right ascension. \\
If `zai' is instead accompanied by a somewhat descriptive statement, it can denote in principle a 
very precise position in connection with the asterism (like that in BC 48).

\begin{figure}
\includegraphics[angle=270,width=0.5\textwidth,trim=0 55 0 0]{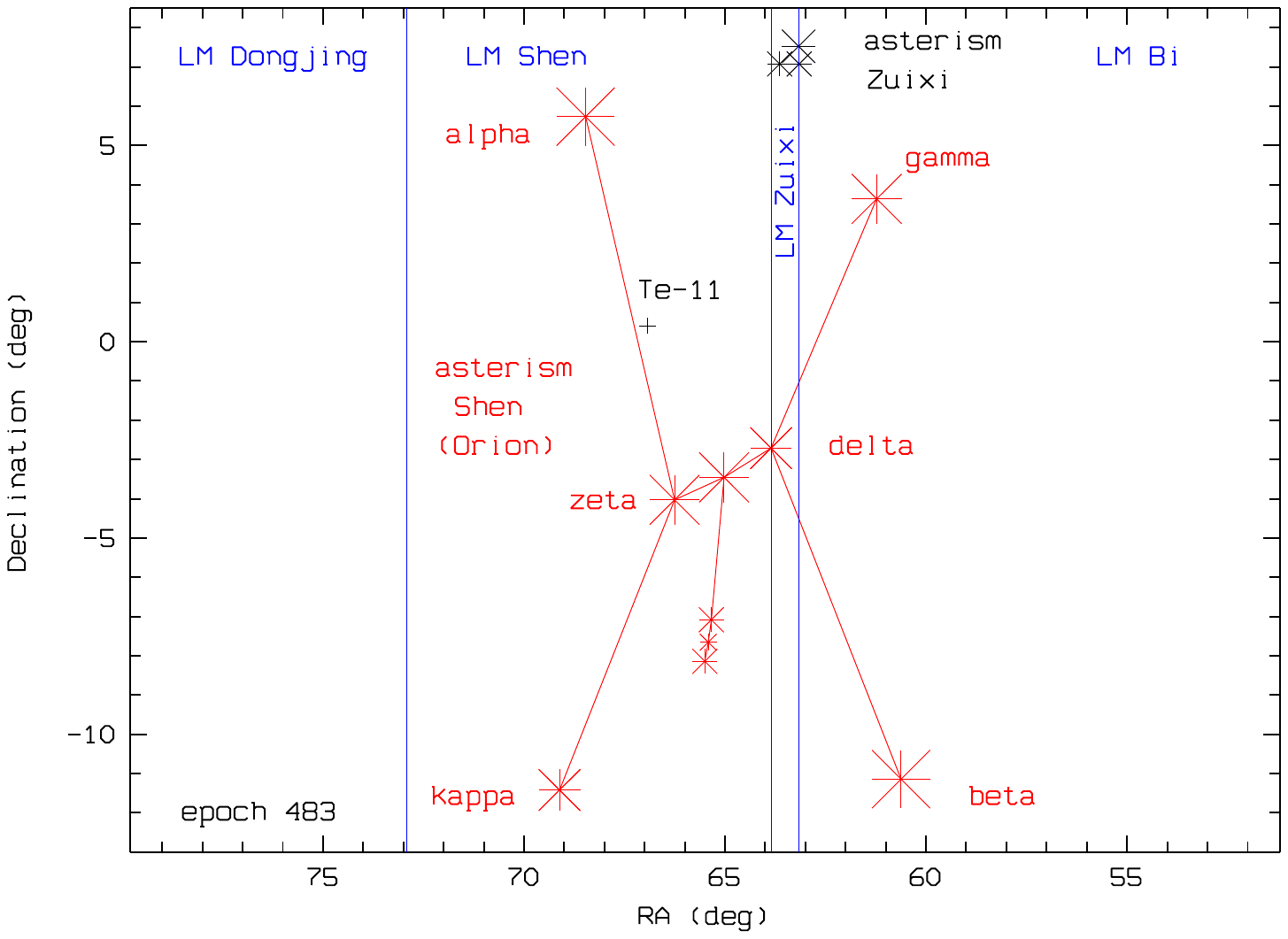}
\caption{The Chinese 
lunar mansion asterisms Shen and Zuixi in Orion in red and black, respectively, plus Te-11 in black; 
the western borders of the lunar mansion RA ranges (LM)
Zuixi, Shen, and Dongjing in blue; 
Te-11 is in the eastern part of the asterism Shen.}
\end{figure}

The wording `zai Shen dong' points to a position given in a descriptive manner: 
If it were not shortened, then it could mean a location 
east 
of (i.e. outside) asterism Shen 
(not necessarily limited by the LM range of Shen) or in/at Shen's 
east 
(Fig. B.1).
However, such rough statements are usually only used to describe 
positions of asterisms relative to others (see Pan 1989, Sun \& Kistemaker 1997).
Alternatively, the wording `zai Shen dong' can point to positions of short-term, 
highly dynamic transients (e.g. bolides), so that the position could not be well constrained. 

The ways to localize a specific star or a more static phenomenon are obviously more sophisticated: 
If one supposes an originally longer version, e.g. similar to that in BC 48, it would be conceivable 
that the concatenated positional statement `zai Shen dong' reflects a specific location in Shen's 
east,
i.e. it could denote the position of Te-11.

The context, see (b), points to a very rough localization. The positional statement for AD 483 remains very imprecise.

(b) Object type:
Stephenson \& Green (2009) `strongly suggest that it was a comet'. 
Miszalski et al. (2016) doubt the comet interpretation, but followed Nickiforov (2010), 
whose methods are unsatisfactory (Neuh\"auser \& Neuh\"auser 2021, therein footnote 2). 
Yet, an hour angle of about 9-11h from the Sun (Shen 
east mid-November to mid-December) 
is not in favor for a cometary sighting.
And, indeed, the description does not seem to report simply a comet, but a phenomenon somewhat similar:  \\
(1) the term `ke xing' is not used exclusively for star-like objects (even though it is the best term to qualify novae); \\
(2) the form/size of the transient is of irregular extension (`like a dipper'); \\
(3) The substantive `bo' used here means `radiance';
in contrast, the compound `xingbo', often translated as `fuzzy star' (or `comet'), in fact literally means 
`star to become fuzzy', it is mostly found in cometary observations (see Pankenier et al. 2008, p. 6), 
but here there is `bo' without `xing' (`star'); \\ 
(4) the position could originally be reported only roughly (understandable for a fast moving object); and \\ 
(5) there is no information about the duration, which is often given for comets. 

Hence, a bolide or meteor might be a credible option. 
Even more, the record is embedded in a larger context (see Appendix E for source) -- two events during daylight precede, 
which sound like observations of bolides or meteors: \\
`Year 10, month 8 (486 Sep 14 to Oct 13) [day missing], chen [double-]hour (corresponds to 7:00-9:00 am). 
There was a star that fell like flowing fire in three streaks (dao). 
Wuyin day ([15]; i.e. Oct 3, assuming year 10, month 8, AD 486): again there was a flowing star that emerged [from] 
a place one zhang SW of the sun, flowing NW, large as Supreme White (Venus).
Reaching south [i.e. `wu', which may be a scribal error for `ren', i.e., west-of-north, 
both within the 24-point-compass], [and then] west [it] broke into two pieces, tail length five chi, 
and again divided in two, entering the space of the clouds. 
[If it] continued [to be] seen, and the affair [went] on and on, 
[would] later generations follow on its heels, reaching the point of dividing and collapsing?' 

Then, the author or redactor recalls the event discussed here for AD 483 which happened three years earlier: \\
`Before this, year 7, month 10 (AD 483), there was a guest star, 
form [or: size] like a dipper (dou), it was located in/at Shen['s] east 
(zai Shen dong), similar to a `bo'. 
The prognostication said: "[When there is] a great minister who seizes the charge of the lord, 
then the harvest is early and grain is expensive."' \\
After this recollection, the text continues with a further event in `year 10' (AD 486), 
now reporting a planetary conjunction of Mars and Jupiter 
(indeed, a close conjunction of Mars and Jupiter happened in early Nov 486).

The record's description of the phenomenon, 
`form [or: size] like 
a dou' -- it is the same graphic `dou' as in Nandou (see Fig. A.1), 
looking like a peck measure or dipper -- indicates an irregularly extended source with a brightening or appearance
`similar to a bo'. The form of a dipper may be compared to a bolide
with a large extension on one end (bowl/head/box) and a long, narrow structure on the other end (handle/tail). 

There is no information left in the record on the exact date, the timing during the night,
nor the duration, motion, or dynamics.

In sum, the link between Te-11 and the transient of AD 483 is not plausible. 
Most probably, the transient mentioned for AD 483 was a bolide or large meteor.

\section{Details on the AD 1645 record}

(a) Position: 
The wording `entered Yugui' signals 
a position within the 
lunar mansion asterism
of Yugui -- a small box of four stars 
($\theta, \eta, \gamma, \zeta$ Cnc) enclosing M44; 
alternatively, a position within the 
lunar mansion RA range 
of Yugui could be meant. 
Yet, AT Cnc (right ascension RA=8h 7m at epoch 1645) is located about $5^{\circ}$ to the NNW of $\eta$ Cnc 
outside the Yugui box. Also, it lies clearly outside the 
lunar mansion RA range 
(LM) Yugui: 
its RA spans from the undisputed determinative star $\theta$ Cnc (e.g., Pan 1989, Sun \& Kistemaker 1997) 
of Yugui (LM 23) at right ascension 8h 11m to $\delta$ Hya of Liu (LM 24) at 8h 19m (epoch 1645). 
Instead, AT Cnc is in Dongjing (LM 22), which spans from its determinative star $\mu$ Gem (RA=6h 2m at epoch 1645)
to $\theta$ Cnc -- this is not clearly stated in Shara et al. (2017a) nor Hoffmann (2019).
In sum, AT Cnc is not located within Yugui,
irrespective of whether or not the 
lunar mansion RA range or the lunar mansion asterism 
was meant. See Fig. C.1.
That the term `xiu' is not present in the text (used as questionable argument in Shara et al. 2017a and also Hoffmann 2019 
that the asterism might be meant) does not modify the conclusion.

For the nova shell of AT Cnc, already Warner (2016) suggested an association to the Korean record of AD 1645 
due to the very rough positional congruency listed by Nickiforov (2010); 
in the latter, however, the historical records are treated in an unsatisfactory manner (see Neuh\"auser \& Neuh\"auser 2021, therein footnote 2) 
and questionable conclusions are drawn (e.g., that phenomena called `fuzzy' would be credible nova candidates
is clearly against the consensus as summarized in Pankenier et al. 2008 and Neuh\"auser \& Neuh\"auser 2021, therein section 3.1). 

\begin{figure}
\includegraphics[angle=270,width=0.5\textwidth,trim=0 55 0 0]{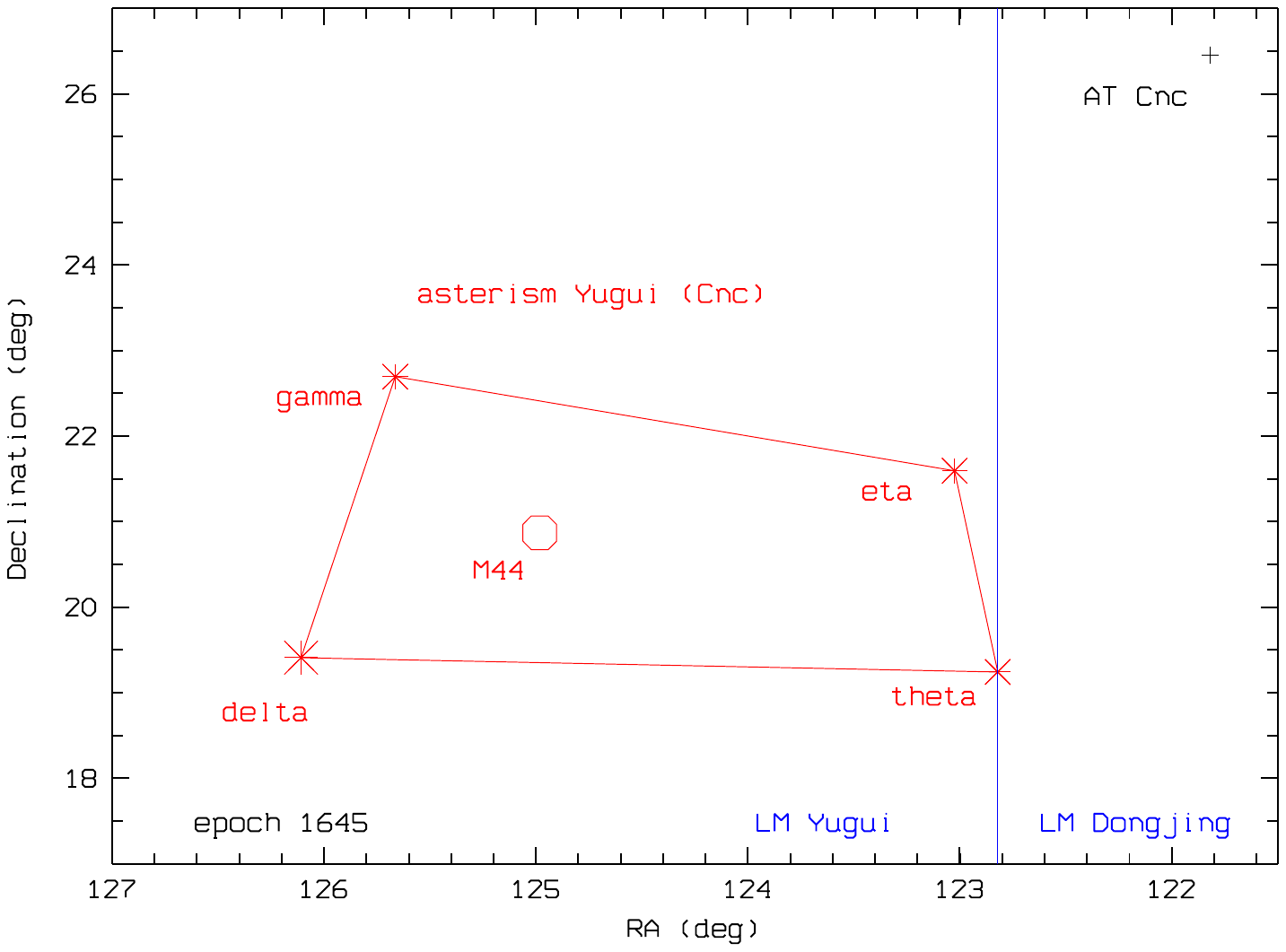}
\caption{The lunar mansion asterism Yugui with the M44 cluster in Cnc in red;
the western border of lunar mansion RA range (LM) Yugui in blue;
AT Cnc is in lunar mansion RA range
LM) Dongjing.}
\end{figure}

(b) Object type:
The use of the otherwise common verb `enter' describes in astronomical reports motion by, 
e.g., moon, planets, or comets. Ho (1966, p. 37) refers to a Chinese standard work of AD 1628:  \\
`the term ju [enter, in Pinyin transcribed as ru] ... is defined as “to move and get into the midst of 
(a constellation or the moon)”'. \\
Relying on Ho (1966) and their own examination, Stephenson \& Green (2009, p. 35) state: \\
`If a guest star was said to “enter” a particular star group, then it may be reasonably concluded that it was moving'. \\
Thus, the report from Korea for AD 1645 would not mirror a fixed star. 
NB: One Japanese record in the context of SN 1006 has `enter' into a specific asterism (see Xu et al. 2000) -- 
due to its excessive brightness (and/or rapidly changing brightness), 
some observer may have thought that its separation to some nearby star was changing.

The phrase `large star' could indicate brightness and/or extension, for instance, 
SN 1006 is reported like that (Xu et al. 2000), but it is otherwise scarcely found in records of 
guest stars --
the nova of AT Cnc with an expected peak brightness of m=1.3$\pm$1.4 mag (Table 1) might hardly be described as `large'.
 
Interestingly, for 1031 Oct 4, the same Korean work has exactly the same phrase as in AD 1645 (Xu et al. 2000) --
that day, Mars `entered' the Yugui asterism box close to $\eta$ Cnc. 
(Warner 2016 mentioned this record together with AD 1645 as nova candidates for AT Cnc.)
The first character in a common name for 
Mars
(`huo' in `huo xing' for `fire star') 
is quite similar to the character for `large' (`da') (with the latter having a single-horizontal stroke 
rather than two short strokes around the center of the character),
and a scribal error would not be far-fetched -- as also suggested by Park et al. (2024)
for the event in AD 1031.
In AD 1645, Mars `entered' the Yugui box around May 16, about two lunar months later than given 
-- the AD 1645 text has only the month (therefore, it is not listed in Park et al. 2024); 
it was noticed before that this Korean source has deficits (at least) in dating 
(e.g. 
Shara et al. 2017a, therein appendix B).

In general, it can be supposed that Korean court astronomers are sufficiently experienced to identify planet Mars,
see Xu et al. (2000) with observations of clusters of multiple planets in the Yijo Sillok, the Korean standard chronicle,
e.g. on 1646 June 18 (p. 269).

However, the transient in AD 1645 is only transmitted in "Supplemental Documents for Reference" (of the Chŭngbo munhŏn pigo,
see Appendix E) --
its context shows a somewhat lower level: \\
`Supplement: Injo (Chinese Renzu), year 3 (AD 1625), autumn (Aug 3 to Oct 30, dates assume parity with Chinese calendar, 
following Fang 1987): A monstrous star in the eastern direction. Year 4 (AD 1626), month 7 (Aug 22 to Sep 19): 
A monstrous star appeared in the kun direction (SW). \\
Year 5 (AD 1627), spring (Feb 16 to May 14): A monstrous star appeared for several months and did not vanish. \\
Month 9 (Oct 9 to Nov 8): It was also like this. \\
Year 23 (AD 1645), month 2 (Feb 26 to Mar 27): A large star entered Yugui.'  

For the years 1625, 1626, and 1627, the text would be plausible with sightings of planet Venus -- 
but a scribal error for `monstrous star' instead of Venus can be ruled out; 
yet, the bright planet Venus could be called `monstrous', especially given its frequently inauspicious associations. 
The phrase `large star' is not necessarily a scribal error for the `fire star' (huo xing), 
i.e.
planet Mars, regarding the context, planet Mars could be just reported 
as a `large star' -- 
and since Mars was at about $-1.5$ mag on both dates in AD 1645 and 1031, such a description would be reasonable. 
  
In sum, the transmission of AD 1645 is not suitable to establish a link to AT Cnc, 
or even to derive theoretical constraints on binary or nova shell evolution. 

\section{Detailed nova shell expansion model}

The nova shell model calculations were performed according to the theory presented in Taylor (1950),
Ostriker \& McKee (1988), Shu (1992), Truelove \& McKee (1999), and Padmanabhan (2001) consisting of the following three phases over time.

\begin{figure*}
\sidecaption
\includegraphics[angle=0,width=10cm]{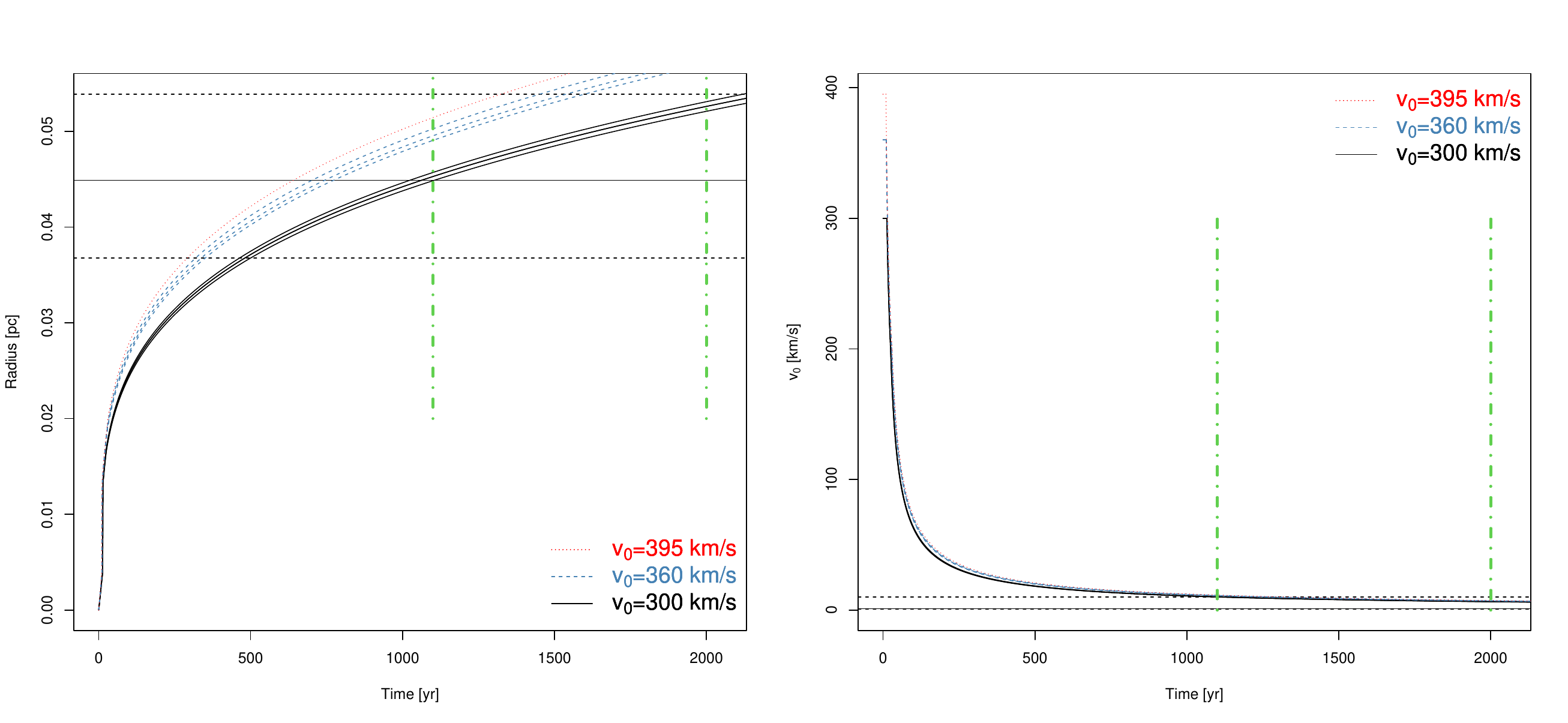}
\caption{Expansion of Te-11 as nova shell:
For 300-395 km s$^{-1}$ initial expansion velocity and an ISM density of 1 cm$^{-3}$,
the current radius 
(0.045 pc) 
and expansion velocity upper limit (10 km s$^{-1}$) can be reached simultaneously within
ca.\,1100 to 2000 yr (green dash-dotted lines);
the two horizontal lines on the right are for 0 and 10 km s$^{-1}$,
as observed by Miszalski et al. (2016): $\le 10$ km s$^{-1}$
for epoch 2012.}
\end{figure*}

(i) Free Expansion (ejecta-dominated phase): The shell of swept-up material in front of the shock does not represent a significant increase 
in mass of the system mass, i.e. the ISM mass previously within the swept-up sphere of radius $r$ is still small 
compared to the ejecta mass (for ISM density $\rho$): 
\begin{equation}
\frac{4 \cdot \pi}{3} \cdot \rho \cdot r^{3} << M_{\rm ejecta}
\end{equation}
As long as the swept-up mass is much smaller than the ejecta mass, 
the velocity $v$ of the shock front remains constant (initial velocity being $v_{0}$), 
the radius of the shell increases linearly over time: 
\begin{equation}
r(t) \sim v_{0} \cdot t
\end{equation}
After free expansion, the Sedov-Taylor phase starts; 
although, in reality, the transition is continuous, a sudden transition should be a good approximation.

(ii) Sedov-Taylor Phase: It is normally assumed that this phase starts when the swept-up mass equals the
ejected mass. The dynamics of the shell can be described by the location of the shock front versus time 
(a self-similar solution), which is determined by the initial energy of the explosion $E$. 
The quantity $E/\rho \cdot t^{2}/r^{5}$ describes the dynamics of the expansion. 
With $k$ as dimensional constant depending in the adiabatic index, the solution requires
\begin{equation}
r(t) = k \cdot (E/\rho)^{1/5} \cdot t^{2/5}~~~~~$\mbox{and}$~~~~~v(t) = 0.4 \cdot r/t.
\end{equation}
This solution describes the expansion of a SNR or nova shell quite well.

(iii) Radiative Cooling 
(snowplough): 
Eventually, the shock slows down, gas is heated less.
When half of the energy has been radiated away, the adiabatic phase is considered to end. 
Typically, the shock speed then drops below ca.\,20 km s$^{-1}$ (with dependence on initial energy and ISM density). 
Most of the material is then swept-up into a dense, cool shell. 
Matter behind the shock cools quickly, pressure is no longer important, 
and the shell moves with constant momentum: 
\begin{equation}
(4 \pi/3) \cdot R^{3} \cdot \rho \cdot v = constant
\end{equation}

Transition times: From relations in Blondin et al. (1998), one can estimate rough transition times.
In Figs. D-1 and D-2, a constant expansion velocity (i.e. no deceleration in the free expansion phase) is plotted
from the explosion until the Sedov-Taylor velocity equals the initial expansion velocity.
Disappearance: When the shock velocity drops to below ca.\,20 km s$^{-1}$, the expansion becomes subsonic and the shell merges with the ISM.

In reality, the situation is more complicated (Truelove \& McKee 1999), e.g. 
that the shocked part of the ISM in front of the ejecta is moving at a different velocity.
Also, the initial ejecta can develop some asymmetry due to WD rotation and magnetic field as well as effects
involving the accretion disk, the donor, possibly ongoing mass flow, and circumbinary medium.
While the Sedov-Taylor formalism was developed for different phenomena,
it has been applied for supernova and also nova remnants to roughly estimate their ages, 
e.g. Shara et al. (2017a) and Santamaria et al. (2019).

In order to find the best values of the initial parameters of a supersonically expanding nova shell (blast wave model)
of Te-11 and AT Cnc, a huge number of simulations is performed for large ranges of initial values.
The results are shown in Figs. 2 (AT Cnc) and D-1 (Te-11). The parameters used for AT Cnc were already
given in Sects. 4 \& 5.

For Te-11, the following parameters are used for the expansion model calculations: current extension of the extended feature 
(semi-major axis) up to 17.6$^{\prime \prime}$ at a distance of $526^{+48}_{-47}$ pc (Gaia DR3),
current expansion velocity $\le 10$ km s$^{-1}$ as upper limit (both from Miszalski et al. 2016),
ejected mass $10^{-6}$~M$_{\odot}$ (from Yaron et al. 2005 for mass and temperature of the WD in Te-11 according to Miszalski et al. 2016)
with $\pm 10\%$ uncertainty, and an ISM density of 0.1 to 1 particle cm$^{-3}$.
All solutions reaching the very low current velocity have a relatively high ISM density (1 cm$^{-3}$)
and a low initial expansion velocity (300-395 km s$^{-1}$).
This results in an age range of 
ca.\,1100 to 2000 yr (before the 2012 epoch of Miszalski et al. 2016). See Fig. D.1.
While this does include the year AD 483, position and object type of the transient observed in that year 
do not fit for a classical nova eruption, see Sect. 3 and Appendix B.
The large age range obtained for the extended feature around Te-11 
also shows how important a reliable historical observation would be to constrain the age.

\begin{figure*}
\sidecaption
\includegraphics[angle=0,width=10cm]{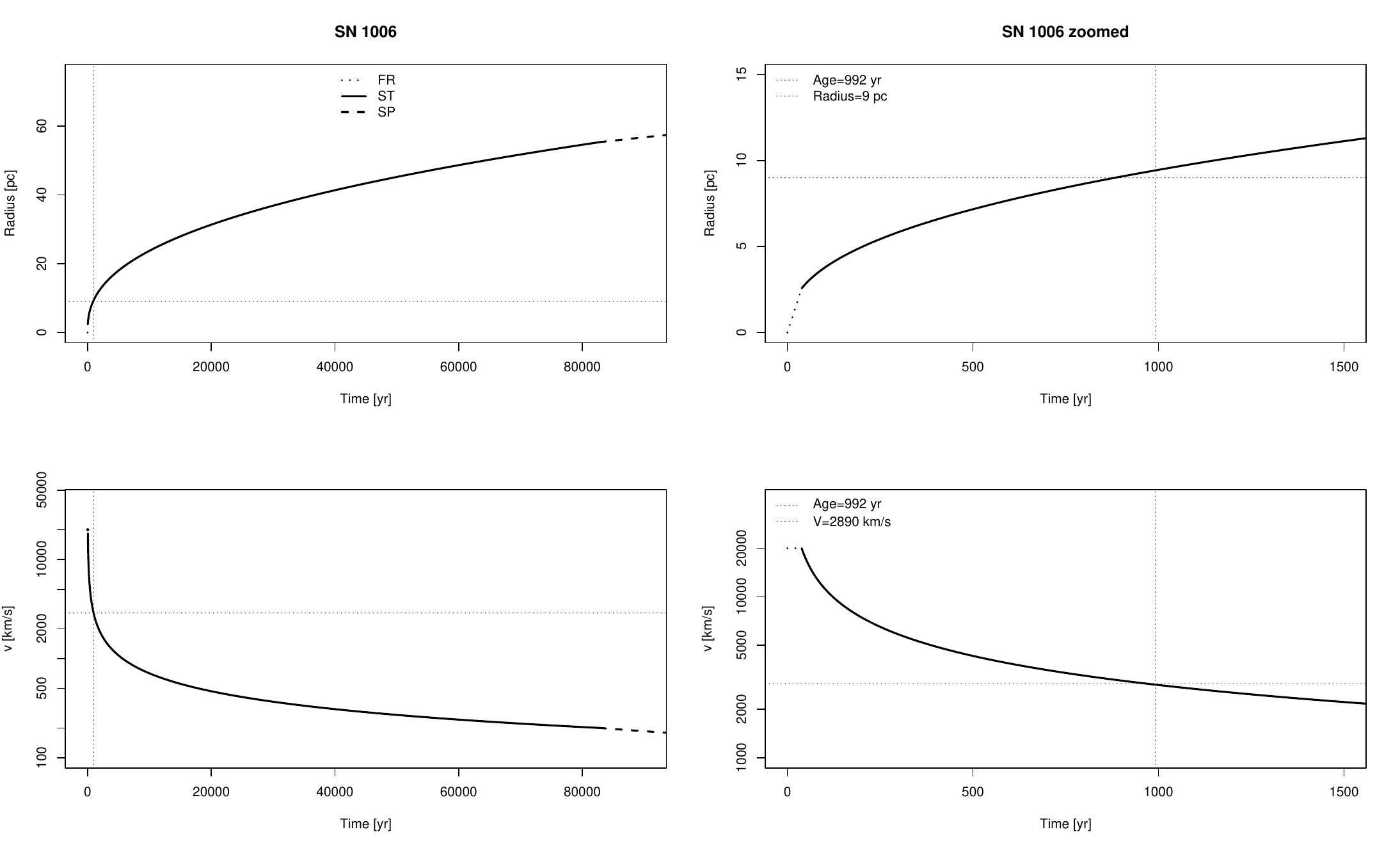}
\caption{Blast wave model calculations for the expansion and deceleration of the SNR of SN 1006:
Radius (upper panels) and expansion velocity $v$ (lower panels, logarithmic scale) are plotted versus time 
with a zoom around its current size (ca.\,9 pc), velocity (2890 km s$^{-1}$), and age (992 yr) on the right;
these current values are indicated by dashed lines (age 992 yr since the observing epoch in Winkler et al. 2003).
Shown are free expansion (FR, dotted lines, first decades), Sedov-Taylor (ST, full lines), 
and 
snowplough 
(SP, broken lines, towards the end of the plotted time range) phases;
the short free expansion phase at the start is hardly resolvable here.
The expansion can be reconstructed well with standard SN Ia input parameters,
also for normal ISM densities (see Sect. 5); see text for the input parameters used.}
\end{figure*}

Extrapolating a decay rate of 10$\pm$3 mmag yr$^{-1}$ obtained for novae from the last century (Duerbeck 1992), 
we would expect that Te-11 (expected peak as nova at m=2.6$\pm$1.4 mag, now at 14.8-20.0 mag, Table 1) 
would have exploded 1220$\pm$315 to 1740$\pm$426 yr before 2012, i.e. between ca.\,BC 155 and AD 1107.
However, as noted before, such a large extrapolation may not be allowed.

NB: The blast wave model has been tested and verified with the remnant of the 
well-known and well-dated historical SN 1006, 
where the parameters are known best.
The age of this brightest and most nearby Galactic supernova within the last two millennia is given with
exceedingly high accuracy and precision due to the historical observations.
SN 1006 was clearly of type Ia (no compact object nor donor is known), 
so that the ejected mass should be 1.3-1.5~M$_{\odot}$, 
irrespective of whether it was double- or single-degenerate. 
The initial ejection velocity for supernovae is known to be in the range of 10000 to 30000 km s$^{-1}$; 
as energy, $10^{51}$ erg is used, 
which is typical for supernovae. From the measurement of the current expansion velocity 
both in radial direction (2890$\pm$100 km s$^{-1}$, Ghavamian et al. 2002) and on sky (280$\pm$8 mas yr$^{-1}$, Winkler et al. 2003),
the distance could be determined to 2.18$\pm$0.08 kpc (Winkler et al. 2003); 
these measurements were performed at the north-western rim of the SNR. 
The current projected radius is 13.6$^{\prime}$, a range of 8.1-9.9 pc is used in the calculation. 
The time since the observations by Winkler et al. (2003) is 992 years, which is used as age of the SNR. 
Hence, all parameters needed for the model calculations are known. 
(As for the nova shells, 
ISM densities of 0.1 to 1 particle cm$^{-3}$ are used.)

As can be seen in Fig. D.2, the model can confirm that the SNR of SN 1006 can reach its current size and velocity 
with the standard blast-wave theory that was also used for Te-11 and AT Cnc: 
there are many solutions for ISM densities around SN 1006 of 0.1 to 0.43 cm$^{-3}$.
The SNR is in the Sedov-Taylor phase. Indeed, the ISM density at the 
north-western rim was otherwise measured to be 0.15-0.25 particles cm$^{-3}$ (Acero et al. 2007).
The blast wave model used here is therefore confirmed.

If the age is considered as free parameter, an age range of 812 to 1062 yr is obtained.
This relatively large range shows again how important an historical observation is to fix the age precisely.

\section{Background on Chinese texts}

Above, in Sects. 2-4, new literal translations for the cases studied were presented
from the original Classical Chinese records from China or Korea.

BC 48.
This record was included in the Han shu (History of the [Western] Han)
“Treatise on Celestial Patterns” (Tianwen zhi), compiled by Ban Zhao (ca.\,AD 48 to 116), 
with the assistance of the brothers Ma Rong (AD 79-166) and Ma Xu (ca.\,79 to after 141), 
some years after the death of her brother Ban Gu (AD 32-92), who had been the primary compiler of the history as a whole.  
While large parts of the treatise are borrowed from Sima Qian's (ca.\,BC 145 to 86) “Tianguan shu”
(Treatise on the Celestial Offices) in his Shiji (Records of the Senior Archivist), 
Ban's work includes a chronicle of aberrant celestial signs through the end of the Western Han (BC 206 to AD 9).  
The “Treatise on Celestial Patterns”, an Eastern Han text (AD 24-220), records (and in some cases analyses) 
celestial signs that indicated the relative health or decay of the previous dynasty.   

AD 483.
This record is included in the astronomical treatise for the Wei shu (History of the [Northern] Wei), 
compiled by Wei Shou (AD 506-572) soon after the conclusion of the dynasty, over a four-year period ending in 554 (Wilkinson 2000).

AD 1645.
This record is included in `References to Celestial Phenomena' (Pinyin `Xiang wei kao') section of the encyclopedic `Chŭngbo munhŏn 
pigo' or `Supplemental Documents for Reference' compiled by the Hongmungwan, a bureau of the Korean government, ca.\,1908.
The transliteration of the Classical Chinese characters to Pinyin follows the Chinese pronunciation.

\end{appendix}


\begin{thebibliography}{}

\bibitem[()]{} Acero, F., Ballet, J., \& Decourchelle, A. 2007, A\&A, 475, 883

\bibitem[()]{} Adam, C. \& Mugrauer, M. 2014, MNRAS, 444, 3459

\bibitem[()]{} Bailer-Jones, C., Rybizki, J., \& Fouesneau, M., et al. 2021, AJ, 161, 147 

\bibitem[()]{} Blondin, J., Wright, E., Borkowski, K., \& Reynolds, S. 1998, ApJ, 500, 342

\bibitem[()]{} Canbay, R., Bilir, S., \"Ozd\"onmez, A., \& Ak, T. 2023, AJ, 165, 163

\bibitem[()]{} Chapman, J. 2021, `Celestial Signs in Three Historical Treatises', in: Csikszentmihalyi, M., Nylan, M. (Eds.), 
Technical Arts in the Han Histories: Tables and Treatises in the Shiji and Hanshu, 
SUNY Press, Albany, pp. 181-211

\bibitem[()]{} Corradi, R., Garcia-Rojas, J., Jones, D., \& Rodriguez-Gil, P. 2015, ApJ, 803, 99

\bibitem[()]{} Della Valle, M. \& Izzo, L. 2020, A\&A Rev., 28, 3

\bibitem[()]{} Downes, R.A. \& Duerbeck, H.W. 2000, AJ, 120, 2007

\bibitem[()]{} Duerbeck, H.W. 1987, ApSS, 131, 461

\bibitem[()]{} Duerbeck, H.W. 1992, MNRAS, 258, 629

\bibitem[()]{} Fang, Z.M. 1987, Zhongguo shi liri he zhongxi liri duizhao biao, 
Shanghai cishu chubanshe, Shanghai

\bibitem[()]{} Frew, D.J., Parker, Q.A., \& Bojicic, I.S. 2016, MNRAS, 455, 1459 

\bibitem[()]{} Ghavamian, P., Winkler, P.F., Raymond, J.C., \& Long, K.S. 2002, ApJ, 572, 888

\bibitem[()]{} G\"ottgens, F., Weilbacher, P.M., Roth, M.M., et al. 2019, A\&A, 626, A69

\bibitem[()]{} Guerrero, M.A., Santamaria, E., Liberato, G., et al. 2025, A\&A, 694, A105

\bibitem[()]{} Ho, P.Y. 1962, Vistas, 5, 127  

\bibitem[()]{} Ho, P.Y. 1966, Astronomical Chapters of Chin Shu, PhD, U Malaysia

\bibitem[()]{} Hoffmann, S.M. 2019, MNRAS, 490, 4194

\bibitem[()]{} Hsi, T. 1957, Smithsonian Contributions to Astrophysics, 2, 109 

\bibitem[()]{} Kiang, T. 1972a, MmRAS, 76, 27

\bibitem[()]{} Kiang, T. 1972b, MNRAS, 157, 477

\bibitem[()]{} Kovetz, A., Prialnik, D., \& Shara, M.M. 1988, ApJ, 325, 828

\bibitem[()]{} Liang, S.Q. 1993, Far East Chinese-English Dictionary, Far East Book, Taipei

\bibitem[()]{} Livio, M., Mazzali, P. 2018, PhR, 736, 1

\bibitem[()]{} Liu 1986, Acta Astronomica Sinica, 27, 276

\bibitem[()]{} Loewe, M. 1994, `The Han View of Comets', 
in: Divination, Mythology and Monarchy in Han China, Cambridge University Press, Cambridge, pp. 61-84

\bibitem[()]{} Lombardi, M., Bouy, H., Alves, J., \& Lada, C.J. 2014, A\&A, 566, A45

\bibitem[()]{} Miszalski, B., Woudt, P.A., Littlefair, P S., et al. 2016, MNRAS, 456, 633 (M+16)

\bibitem[()]{} Monaco, L., Pancino, E., Ferraro, F.R., \& Bellazzini, M. 2004, MNRAS, 349, 1278

\bibitem[()]{} Morrison, L.V. \& Stephenson, F.R. 2004, JHA, 35, 327

\bibitem[()]{} Needham, J. \& Wang, L. 1959, Science and Civilization in China, Vol. 3, 
Mathematics and the Sciences of the Heavens and the Earth, Cambridge University Press, New York

\bibitem[()]{} Neuh\"auser, D.L., Neuh\"auser, R., Mugrauer, M., Harrak, A., \& Chapman, J. 2021b, Icarus, 364, 114278

\bibitem[()]{} Neuh\"auser, R., Neuh\"auser, D.L. 2021, AN, 34, 675

\bibitem[()]{} Neuh\"auser, R., Neuh\"auser, D.L., \& Posch, T. 2020, Terra-Astronomy -- Understanding historical observations to study transient phenomena, 
in: Lago T. (Ed.) Astronomy in Focus -- Proc. Focus Meetings IAU GA (held August 2018, Vienna, Austria), Cambridge University Press, Cambridge, pp. 145-147 

\bibitem[()]{} Neuh\"auser, R., Neuh\"auser, D.L., \& Chapman, J. 2021a, MNRAS, 501, L1 

\bibitem[()]{} Neuh\"auser, R., Torres, G., Mugrauer, M., Neuh\"auser, D.L., Chapman, J., Luge, D., \& Cosci, M. 2022, MNRAS, 516, 693

\bibitem[()]{} Nickiforov, M. 2010, Bulg. Astron. J., 13, 16

\bibitem[()]{} Nogami, D., Masuda, S., Kato, T., \& Hirata, R. 1999, PASJ, 51, 115

\bibitem[()]{} Ostriker, J.P. \& McKee, C.F. 1988, Reviews of Modern Physics, 60, 1

\bibitem[()]{} Padmanabhan, T. 2001, Theoretical Astrophysics, Vol. 2, Stars and Stellar Systems, Cambridge University Press, Cambridge

\bibitem[()]{} Pan, N. 1989, Zhongguo hengxing guance shi (History on Stellar Observations in China), Xue Lin Press, Shanghai

\bibitem[()]{} Pankenier, D.W., Xu, Z., \& Jiang, Y. 2008, Archeoastronomy in East Asia, New York, Cambria

\bibitem[()]{} Park, J., Jeon, J., \& An, H. 2024, AN, 345, e240068

\bibitem[()]{} Raphals, L. 2013, Divination and Prediction in Early China and Ancient Greece, Cambridge University Press, Cambridge 

\bibitem[()]{} Samus, N., Kravtsov, V., Pavlov, M., et al. 1995, A\&AS, 109, 487

\bibitem[()]{} Schaefer, B.E. 2010, ApJS, 187, 275

\bibitem[()]{} Schaefer, B.E. 2018, MNRAS, 481, 3033

\bibitem[()]{} Schaefer, B.E. 2023, MNRAS, 525, 785

\bibitem[()]{} Santamaria, E., Guerrero, M., Ramos-Larios, G., et al. 2019, MNRAS, 483, 3773

\bibitem[()]{} Santamaria, E., Guerrero, M., Ramos-Larios, G., et al. 2020, ApJ, 892, 60

\bibitem[()]{} Seitter, W.C. 1990, Optical Studies of Classical Novae in Outburst, LNP, 369, 79

\bibitem[()]{} Shafter, A.W. 2017, ApJ, 834, 196

\bibitem[()]{} Shara, M.M., Livio, M., Moffat, A., \& Orio, M. 1986, ApJ, 311, 3017 

\bibitem[()]{} Shara, M.M., Martin, C., Seibert, M., et al. 2007, Nature, 446, 159 

\bibitem[()]{} Shara, M.M., Mizusawa, T., Zurek, D., et al. 2012a, ApJ, 756, 107 

\bibitem[()]{} Shara, M.M., Mizusawa, T., Wehinger, P., et al. 2012b, ApJ, 758, 121 (S+12b)

\bibitem[()]{} Shara, M.M., Drissen, L., Martin, T., Alaire, A., Stephenson, F.R. 2017a, MNRAS, 465, 739 (S+17a)

\bibitem[()]{} Shara, M.M., Ilkiewicz, K., Mikolajewska, J., et al. 2017b, Nature, 548, 558 

\bibitem[()]{} Shara, M.M., Lanzetta, K. M., Garland, J. T., et al. 2024, MNRAS, 529, 212 

\bibitem[()]{} Shu, F.H. 1992, Physics of Astrophysics II, Univ. Science Books, Sausalito

\bibitem[()]{} Stephenson, F.R. 1994, Chinese and Korean Star Maps and Catalogs, in: J. B. Harley, D. Woodward (Eds.), 
The History of Cartography, Vol. 2, Book 2,
University of Chicago Press, Chicago and London, pp. 511-578 

\bibitem[()]{} Stephenson, F.R. \& Green, D. 2009, JHA, 40, 31  

\bibitem[()]{} Sun, X. \& Kistemaker, J. 1997, The Chinese sky during the Han: Constellating the stars and society, Brill, Leiden
	
\bibitem[()]{} Tappert, C., Vogt, N., Ederoclite, A., et al. 2020, A\&A, 641, 122 

\bibitem[()]{} Taylor, G. 1950, Proc. R. Soc. A, 101, 159

\bibitem[()]{} Truelove, J.K. \& McKee, C.F. 1999, ApJS, 120, 299

\bibitem[()]{} Van den Bergh, S., Younger, P.F. 1987, A\&AS, 70, 125

\bibitem[()]{} Wang, L. 1999, Gudai Hanyu, Zhonghua shuju, Beijing 

\bibitem[()]{} Wang, Z.R., Liu, J.Y., Gorenstein, P., \& Zombeck, M.V. 1986, HiA, 7, 583

\bibitem[()]{} Wang, Z.R., Li, M., \& Zhao, Y. 2006, CJAA, 6, 625

\bibitem[()]{} Warner, B. 1995, Cataclysmic variable stars, Cambridge University Press

\bibitem[()]{} Warner, B. 2016, Celestial optical transients from 532 BCE to 2015 AD. From the ridiculous to the sublime? 
In: Proceedings of Science, 3rd annual conference on high energy astrophysics in Southern Africa (HEASA2015) 

\bibitem[()]{} Weidmann, W.A., Mari, M.B., Schmidt, E.O., et al. 2020, A\&A, 640, A10

\bibitem[()]{} Wesson, R., Barlow, M.J., Corradi, R.L.M., et al. 2008, ApJ, 688, L21 

\bibitem[()]{} Wilkinson, E. 2000, Chinese History: A Manual, Revised and Enlarged Edition, Harvard University Press, Cambridge

\bibitem[()]{} Williams, R. 2013, AJ, 146, 55

\bibitem[()]{} Winkler, P.F., Gupta, G., \& Long, K.S. 2003, ApJ, 585, 324

\bibitem[()]{} Xu, Z., Jiang, Y., \& Pankenier, D.W. 2000, East Asian Archaeoastronomy: 
historical records of astronomical observations of China, Japan, and Korea, Gordon and Breach, Amsterdam

\bibitem[()]{} Yaron, O., Prialnik, D., Shara, M., \& Kovetz, A. 2005, ApJ, 623, 398

\bibitem[()]{} Yau, K. 1988, An investigation of some contemporary problems in astronomy and 
astrophysics by way of early astronomical records, PhD, U Durham 

\bibitem[()]{} Yi, S. 1984, Qantian xingtu: 2000.0, Beijing

\bibitem[()]{} Yu, J., Khanna, S., Themessl, N., et al. 2023, ApJ, 265, id. 41

\bibitem[()]{} Yun, L. 1845, Yixiang kaocheng xubian 

\bibitem[()]{} Yun, L. 1983, Yixiang kaocheng, Taiwan shangwu yinshuguan, Taipei 

\end{thebibliography}
\end{document}